\documentclass{ieeeaccess}
\usepackage{cite}
\usepackage{scalerel}
\usepackage{tikz}
\NewSpotColorSpace{PANTONE}
\AddSpotColor{PANTONE} {PANTONE3015C} {PANTONE\SpotSpace 3015\SpotSpace C} {1 0.3 0 0.2}
\SetPageColorSpace{PANTONE}   
\usetikzlibrary{svg.path}
\usepackage{amsmath,amssymb,amsfonts}
\usepackage{graphicx}
\usepackage{textcomp}
\usepackage{multirow}
\usepackage[table,xcdraw]{xcolor}
\usepackage{colortbl} 
\usepackage{lipsum}
\usepackage{algorithm}
\usepackage{algpseudocode}
\algrenewcommand\algorithmicrequire{\textbf{Input:}}
\algrenewcommand\algorithmicensure{\textbf{Output:}}
\usepackage{easyReview}
\usepackage{soul}
\usepackage{enumitem}

\usepackage{bm}
\makeatletter
\AtBeginDocument{\DeclareMathVersion{bold}
\SetSymbolFont{operators}{bold}{T1}{times}{b}{n}
\SetSymbolFont{NewLetters}{bold}{T1}{times}{b}{it}
\SetMathAlphabet{\mathrm}{bold}{T1}{times}{b}{n}
\SetMathAlphabet{\mathit}{bold}{T1}{times}{b}{it}
\SetMathAlphabet{\mathbf}{bold}{T1}{times}{b}{n}
\SetMathAlphabet{\mathtt}{bold}{OT1}{pcr}{b}{n}
\SetSymbolFont{symbols}{bold}{OMS}{cmsy}{b}{n}
\renewcommand\boldmath{\@nomath\boldmath\mathversion{bold}}}
\makeatother
\def\BibTeX{{\rm B\kern-.05em{\sc i\kern-.025em b}\kern-.08em
    T\kern-.1667em\lower.7ex\hbox{E}\kern-.125emX}}

\usepackage{xcolor}
\usepackage{pict2e}

\definecolor{orcidlogocol}{HTML}{A6CE39}
\tikzset{
	orcidlogo/.pic={
		\fill[orcidlogocol] svg{M256,128c0,70.7-57.3,128-128,128C57.3,256,0,198.7,0,128C0,57.3,57.3,0,128,0C198.7,0,256,57.3,256,128z};
		\fill[white] svg{M86.3,186.2H70.9V79.1h15.4v48.4V186.2z}
		svg{M108.9,79.1h41.6c39.6,0,57,28.3,57,53.6c0,27.5-21.5,53.6-56.8,53.6h-41.8V79.1z M124.3,172.4h24.5c34.9,0,42.9-26.5,42.9-39.7c0-21.5-13.7-39.7-43.7-39.7h-23.7V172.4z}
		svg{M88.7,56.8c0,5.5-4.5,10.1-10.1,10.1c-5.6,0-10.1-4.6-10.1-10.1c0-5.6,4.5-10.1,10.1-10.1C84.2,46.7,88.7,51.3,88.7,56.8z};
	}
}

\newcommand\orcidicon[1]{\href{https://orcid.org/#1}{\mbox{\scalerel*{
				\begin{tikzpicture}[yscale=-1,transform shape]
					\pic{orcidlogo};
				\end{tikzpicture}
			}{|}}}}

\usepackage{hyperref} 

\hypersetup{
	colorlinks=false,
	linkbordercolor=white,
	urlbordercolor=white,
	pdfborder={0 0 0}
}

\begin{document}
\history{Date of publication xxxx 00, 0000, date of current version xxxx 00, 0000.}
\doi{10.1109/ACCESS.2026.0429000}

\title{Apnea Burden-Guided Framework: Enhancing Out-of-Distribution Generalization in PPG-Based Sleep Apnea Characterization}
\author{\uppercase{Mantas Rinkevi\v{c}ius}\authorrefmark{\orcidicon{0000-0002-0989-3023}1},\
\uppercase{Oskar Pfeffer}\authorrefmark{\orcidicon{0000-0002-9456-7694}2},\
\uppercase{Amal Alissa}\authorrefmark{\orcidicon{0009-0001-9643-0608}2,3},\
\\
\uppercase{Nando Hegemann}\authorrefmark{\orcidicon{0000-0003-3953-9006}2},\ 
\uppercase{and}
\uppercase{Vaidotas Marozas}\authorrefmark{\orcidicon{0000-0002-6879-5845}1,4},\ \IEEEmembership{Member, IEEE}}

\address[1]{Biomedical Engineering Institute, Kaunas University of Technology, LT-51423 Kaunas, Lithuania}
\address[2]{Physikalisch-Technische Bundesanstalt, Bundesallee 100, NDS-38116 Brunswick, Germany}
\address[3]{Freie Universität Berlin, Kaiserswerther str. 16-18, 14195 Berlin, Germany}
\address[4]{Faculty of Electrical and Electronics Engineering, Kaunas University of Technology, LT-51368 Kaunas, Lithuania}
\tfootnote{This work was supported by 22HLT01 QUMPHY. This project (22HLT01 QUMPHY) has received funding from the European Partnership on Metrology, co-financed from the European Union Horizon Europe Research and Innovation Programme and by the Participating States.}

\markboth
{Rinkevi\v{c}ius et al.: AB-Guided Framework: Enhancing OOD Generalization in PPG-Based Sleep Apnea Characterization}
{Rinkevi\v{c}ius et al.: AB-Guided Framework: Enhancing OOD Generalization in PPG-Based Sleep Apnea Characterization}

\corresp{Corresponding author: Mantas Rinkevičius (e-mail: mantas.rinkevicius@ktu.lt)}

\begin{abstract}
(1)~\emph{Background:} Sleep apnea is a common sleep-related breathing disorder associated with substantial cardiovascular and metabolic risk. Although overnight polysomnography remains the reference standard for diagnosis, its complexity and cost limit its suitability for long-term preventive monitoring at home. In wearable systems, arterial blood oxygen saturation is commonly used as the main predictor, whereas additional morphological features of the photoplethysmographic pulse wave are usually underexplored. (2)~\emph{Objective:} This study proposes a novel apnea burden prediction-based framework for sleep apnea severity assessment and investigates the influence of photoplethysmographic features on model performance and out-of-distribution generalization. (3)~\emph{Methods:} The proposed framework first predicts apnea burden as a continuous measure, which is subsequently converted into the clinically relevant apnea-hypopnea index for subject-level classification into four severity groups in both in-distribution and out-of-distribution data. Three artificial neural network architectures were evaluated, and the performance metrics were averaged over five independent runs with different fixed random seeds. (4)~\emph{Results:} During out-of-distribution testing, the combination of photoplethysmographic features and arterial blood oxygen saturation led to increases of approximately 15.72\% in macro-sensitivity, 9.22\% in macro-accuracy, 16.01\% in macro-F1-score, 11.08\% in Cohen's kappa, and 13.22\% in Matthews correlation coefficient, compared with using arterial blood oxygen saturation alone. The low-complexity convolutional-recurrent models achieved the highest overall performance. (5)~\emph{Conclusion:} The results indicated that the proposed apnea burden-guided framework, combined with photoplethysmographic features and arterial blood oxygen saturation, improves sleep apnea characterization while showing encouraging out-of-distribution performance on an independent external dataset. Moreover, simpler hybrid architectures demonstrated strong potential for robust home-based preventive monitoring.
\end{abstract}


\titlepgskip=-21pt

\maketitle

\section{Introduction}
\label{sec:introduction}
\PARstart{S}{leep} apnea is a common sleep-related breathing disorder characterized by recurrent episodes of partial or complete upper airway obstruction during sleep, leading to intermittent hypoxia, sleep fragmentation, daytime sleepiness, and increased cardiovascular risk\textcolor{blue}{\hyperlink{ref1}{\cite{ref1}}},\textcolor{blue}{\hyperlink{ref1b}{\cite{ref1b}}},\textcolor{blue}{\hyperlink{ref1c}{\cite{ref1c}}},\textcolor{blue}{\hyperlink{ref3}{\cite{ref3}}}. Epidemiological studies indicate that a substantial proportion of the adult population is affected (from 3\% to 28\%), with prevalence increasing due to aging and obesity\textcolor{blue}{\hyperlink{ref2}{\cite{ref2}}},\textcolor{blue}{\hyperlink{ref2b}{\cite{ref2b}}}. The "gold standard" for sleep apnea diagnosis is polysomnography (PSG), a comprehensive overnight examination that records multiple physiological signals, including respiratory airflow, electrocardiography (ECG), electroencephalography (EEG), and arterial blood oxygen saturation (SpO2). During a PSG test, sleep apnea severity is clinically quantified using the apnea-hypopnea index (AHI), defined as the average number of apnea (complete cessation) and hypopnea (partial reduction in airflow) events lasting $\geq$10~s per hour of sleep, associated with $\geq$3\% oxygen desaturation\textcolor{blue}{\hyperlink{ref2c}{\cite{ref2c}}},\textcolor{blue}{\hyperlink{ref2d}{\cite{ref2d}}}. According to\textcolor{blue}{\hyperlink{ref2d}{\cite{ref2d}}}, sleep apnea is classified into four categories: normal (AHI $<$ 5), mild (5 $\leq$ AHI $<$ 15), moderate (15 $\leq$ AHI $<$ 30), and severe (AHI $\geq$ 30). Despite the high diagnostic accuracy of PSG in estimating the AHI, it remains costly, complex, and impractical for long-term monitoring in home environments.

\begin{figure*}[!t] 
	\centering 
	\includegraphics[width=15cm]{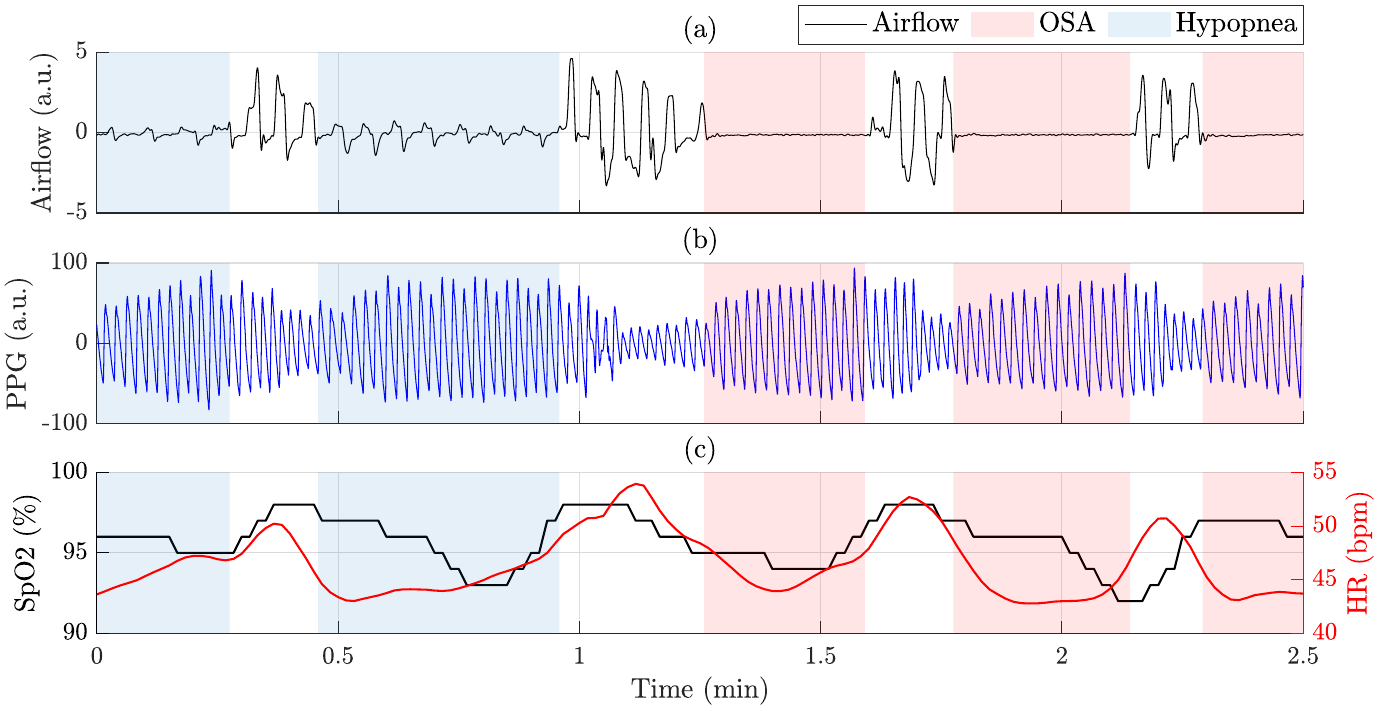} 
	\caption{The changes of signals during obstructive sleep apnea--OSA (complete cessation) and hypopnea (partial reduction in airflow) episodes: (a) respiratory airflow, (b) finger PPG signal, (c) arterial blood oxygen saturation, SpO2, and heart rate, HR.}
	\label{Fig1}
\end{figure*}

To overcome these limitations, wearable-based approaches using photoplethysmography (PPG), which reflects peripheral blood volume variations, and SpO2 have gained increasing attention\textcolor{blue}{\hyperlink{ref4}{\cite{ref4}}},\textcolor{blue}{\hyperlink{ref5}{\cite{ref5}}},\textcolor{blue}{\hyperlink{ref5b}{\cite{ref5b}}}. In current practice, SpO2, derived from dual-wavelength PPG signals (typically red and infrared), is the most commonly used predictor for sleep apnea characterization in wearable devices designed for home-based preventive monitoring. In some cases, heart/pulse rate, derived from a single channel PPG, is also considered as an additional feature. However, the morphological characteristics of the PPG waveform are typically not exploited, and the valuable physiological information contained in pulse wave dynamics remains largely underutilized. Figure~\textcolor{blue}{\ref{Fig1}} illustrates typical changes in respiratory airflow, PPG signal, SpO2, and heart rate during apnea/hypopnea episodes, highlighting that sleep apnea affects multiple physiological factors beyond oxygen desaturation alone. Due to reduced airflow, SpO2 decreases, whereas heart rate typically exhibits a transient reduction during respiratory events, followed by a sudden post-event increase as part of the arousal response, resulting in a U-shaped pattern.

Early studies have demonstrated that PPG signals contain relevant patterns related to cardiovascular and autonomic responses to apnea events, including pulse rate variability and amplitude modulation\textcolor{blue}{\hyperlink{ref6}{\cite{ref6}}},\textcolor{blue}{\hyperlink{ref6b}{\cite{ref6b}}},\textcolor{blue}{\hyperlink{ref7}{\cite{ref7}}},\textcolor{blue}{\hyperlink{ref7b}{\cite{ref7b}}} (see Figure~\textcolor{blue}{\ref{Fig1}b}). For instance, a study\textcolor{blue}{\hyperlink{ref6b}{\cite{ref6b}}} employed a combination of PPG and SpO2 signals for sleep apnea event detection, showing promising capability to distinguish between central and obstructive apnea types, as well as apnea and hypopnea events. Their findings suggest that incorporating PPG-derived features enhances sleep apnea severity assessment beyond conventional SpO2-based detectors.

Recent advances in deep learning have significantly improved the analysis of physiological signals by enabling automatic extraction of complex temporal patterns\textcolor{blue}{\hyperlink{ref8}{\cite{ref8}}},\textcolor{blue}{\hyperlink{ref8b}{\cite{ref8b}}},\textcolor{blue}{\hyperlink{ref8b}{\cite{ref8c}}}. Artificial neural networks, including convolutional and recurrent architectures, have been successfully applied to sleep apnea detection using pulse rate and SpO2 time series\textcolor{blue}{\hyperlink{ref9}{\cite{ref9}}},\textcolor{blue}{\hyperlink{ref10}{\cite{ref10}}}. In particular, hybrid convolutional-recurrent models have demonstrated strong capability in learning both local and long-term temporal features. For instance, the DeepSleepNet model was proposed for automatic sleep stage scoring based on raw single-channel EEG\textcolor{blue}{\hyperlink{ref11}{\cite{ref11}}}, and the ApneaNet architecture was introduced to detect obstructive sleep apnea using ECG signal analysis\textcolor{blue}{\hyperlink{ref13}{\cite{ref13}}}. More recent works have explored advanced architectures, including transformer-based models\textcolor{blue}{\hyperlink{ref12}{\cite{ref12}}},\textcolor{blue}{\hyperlink{ref14}{\cite{ref14}}},\textcolor{blue}{\hyperlink{ref14b}{\cite{ref14b}}},\textcolor{blue}{\hyperlink{ref14c}{\cite{ref14c}}} for processing EEG, ECG or SpO2 raw signals.

However, a critical limitation of most proposed home-oriented approaches based on PPG is their reliance on SpO2 as the primary predictor of sleep apnea\textcolor{blue}{\hyperlink{ref14c}{\cite{ref14c}}},\textcolor{blue}{\hyperlink{ref15}{\cite{ref15}}},\textcolor{blue}{\hyperlink{ref15b}{\cite{ref15b}}},\textcolor{blue}{\hyperlink{ref15c}{\cite{ref15c}}},\textcolor{blue}{\hyperlink{ref15d}{\cite{ref15d}}},\textcolor{blue}{\hyperlink{ref15e}{\cite{ref15e}}}, while overlooking pulse wave morphological features. This may limit the ability to capture more subtle physiological changes in PPG morphology and reduce generalizability when applied to data from different distributions as studies\textcolor{blue}{\hyperlink{ref6b}{\cite{ref6b}}},\textcolor{blue}{\hyperlink{ref14c}{\cite{ref14c}}},\textcolor{blue}{\hyperlink{ref15}{\cite{ref15}}},\textcolor{blue}{\hyperlink{ref15c}{\cite{ref15c}}} performed with only one dataset. In this context, in-distribution (ID) data refers to samples drawn from the same distribution as the training data, whereas out-of-distribution (OOD) data corresponds to samples originating from different populations, devices, or acquisition conditions. Models often perform well on ID data but exhibit performance degradation on OOD datasets. Recent studies have highlighted the importance of addressing this issue. For example, the solely PPG-based ApSense study emphasized the need for multi-dataset validation\textcolor{blue}{\hyperlink{ref16}{\cite{ref16}}}, whereas the SleepPPG-Net2 explicitly addressed cross-dataset generalization for sleep staging from raw PPG time series\textcolor{blue}{\hyperlink{ref17}{\cite{ref17}}}.

In this study, we hypothesize that combining PPG pulse wave features with SpO2 in artificial neural networks improves classification accuracy and enhances OOD generalization. In addition, we propose a couple of novel and underexplored features for sleep apnea assessment. Therefore, this work investigates the influence of PPG-derived features on model performance under distributional shifts. We employ a novel sleep apnea burden (AB) prediction framework and compare three network architectures of varying complexity with different predictor combinations. The AB metric is defined as the proportion of time affected by respiratory events, thus providing a continuous representation of disease severity. While AHI is widely used in clinical practice, AB offers a complementary perspective by capturing the temporal extent of respiratory disturbances. Our proposed approach enables continuous estimation of AB, which is subsequently transformed into the clinically relevant AHI metric, allowing subject-level severity classification, consistent with clinical practice.

The main contributions of this study are summarized as follows:
\begin{enumerate}
	\item A model-agnostic AB regression-based framework is proposed for sleep apnea severity assessment using PPG-derived features and SpO2.
	\item The contribution of different PPG feature combinations, including the novel and underexplored features, is systematically evaluated.
	\item The proposed framework is assessed under both ID and OOD scenarios using three representative deep learning architectures.
	\item The potential advantages of regression-based sleep apnea characterization over conventional classification-based approaches are discussed.
\end{enumerate}

The remainder of this paper is organized as follows. Section~\textcolor{blue}{\ref{Section2}} presents the materials and methods, including the dataset description, data partitioning and preprocessing, PPG feature extraction, and the investigated model architectures. Section~\textcolor{blue}{\ref{Section3}} reports the experimental results, including parameter settings, performance of different model architectures, and comparison of predictor combinations. Section~\textcolor{blue}{\ref{Section4}} discusses the main findings, highlighting the novelty of the proposed method, the impact of model architectures and features, as well as limitations of the study and directions for future work. Finally, Section~\textcolor{blue}{\ref{Section5}} concludes the obtained results.

\section{Materials and methods}
\label{Section2}

\begin{figure}[h] 
	\centering 
	\includegraphics[width=\columnwidth]{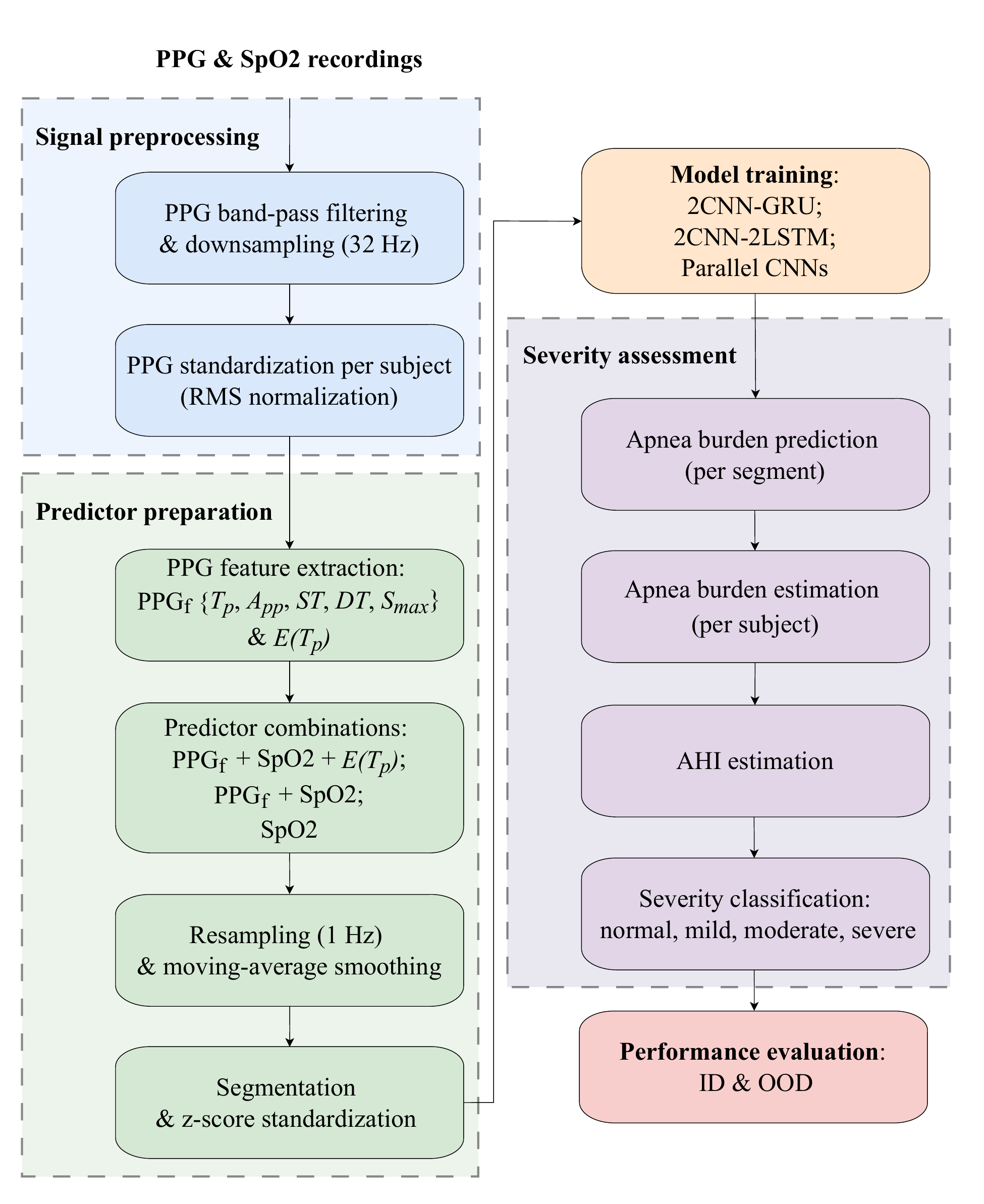}
	\caption{The proposed workflow for the AB-guided framework used to process PPG and SpO2 time series~for~sleep~apnea~severity~assessment. PPG$_f$ corresponds to five pulse wave features, whereas $E(T_p)$ is a new proposed higher-level feature.} 
	\label{Fig_Structure}
\end{figure}

This section describes the datasets used and the research workflow for the proposed AB-guided sleep apnea severity assessment framework. The proposed research workflow consists of (see Figure~\textcolor{blue}{\ref{Fig_Structure}}): (i)~signal preprocessing, (ii)~predictor preparation, (iii)~model training, (iv)~severity assessment, and (v)~performance evaluation. A pseudocode of the proposed AB-guided framework is provided in Algorithm~\textcolor{blue}{\ref{PseudoCode}}.

\begin{algorithm}[!t]
	\caption{Pseudocode of the proposed AB-guided framework}
	\label{PseudoCode}
	\begin{algorithmic}[1]
		
		\Require
		\hspace{5pt}$M$ test subjects; $N_j$ segments for subject $j$
		
		Segmented predictor sequences:
		$\mathcal{X}_{\mathrm{tr}}$, $\mathcal{X}_{\mathrm{val}}$, $\mathcal{X}_{\mathrm{test}}$
		
		Reference segment-level AB values:
		$\mathcal{Y}_{\mathrm{tr}}$, $\mathcal{Y}_{\mathrm{val}}$, $\mathcal{Y}_{\mathrm{test}}$
		
		Training data-derived mean event duration $D_{\mathrm{mean}}$
		
		Reference severity class ${c}_j$
		
		\vspace{10pt}
		
		\Ensure
		Predicted subject-level $\widehat{AB}_j$
		
		\hspace{3pt}Estimated apnea-hypopnea index $\widehat{AHI}_j$
		
		\hspace{3pt}Predicted severity class $\widehat{c}_j$
		
		\vspace{10pt}
		
		\State Estimate the feature-wise z-score normalization parameters
		$\boldsymbol{\mu}_{\mathrm{tr}}$ and
		$\boldsymbol{\sigma}_{\mathrm{tr}}$
		from $\mathcal{X}_{\mathrm{tr}}$
		
		\State Standardize all subsets ($\mathcal{X}_{\mathrm{tr}}$, $\mathcal{X}_{\mathrm{val}}$, $\mathcal{X}_{\mathrm{test}}$) using
		$\boldsymbol{\mu}_{\mathrm{tr}}$ and
		$\boldsymbol{\sigma}_{\mathrm{tr}}$ values
		
		\State Initialize the selected model architecture network $f_{\boldsymbol{\theta}}$
		
		\State Train $f_{\boldsymbol{\theta}}$ using
		$\mathcal{X}_{\mathrm{tr}}$ and $\mathcal{Y}_{\mathrm{tr}}$
		
		\State Select the model parameters
		$\boldsymbol{\theta}^{*}$ according to the validation loss using $\mathcal{X}_{\mathrm{val}}$ and $\mathcal{Y}_{\mathrm{val}}$
		\vspace{10pt}
		\For{$j = 1$ to $M$}
		\vspace{10pt}
		\For{$s = 1$ to $N_j$}
		
		\State
		$\widehat{AB}_{j,s}
		\gets
		f_{\boldsymbol{\theta}^{*}}
		\left(\mathcal{X}^{j,s}_{\mathrm{test}}\right)$
		
		\EndFor
		\vspace{10pt}
		\State
		$\widehat{AB}_{j}
		\gets
		\operatorname{Aggregate}
		\left(
		\widehat{AB}_{j,1},
		\ldots,
		\widehat{AB}_{j,N_j}
		\right)$
		
		\State
		$\widehat{AHI}_{j}
		\gets
		\operatorname{Convert}
		\left(\widehat{AB}_{j},
		D_{\mathrm{mean}}
		\right)$
		according to Equation~\ref{AHI_AB}
		
		\State
		$\widehat{c}_{j}
		\gets
		\operatorname{Classify}
		\left(
		\widehat{AHI}_{j}
		\right)$
		
		\EndFor
		\vspace{10pt}
		\State \Return
		$\left\{
		\widehat{AB}_{j},
		\widehat{AHI}_{j},
		\widehat{c}_{j}
		\right\}_{j=1}^{M}$
		
		\vspace{10pt}
		
		\State Evaluate differences between ${c}_j$ and $\widehat{c}_j$
		
	\end{algorithmic}
\end{algorithm}

\subsection{Dataset description}
\label{Dataset}
\subsubsection{MESA data}	 

In this work, the Multi-Ethnic Study of Atherosclerosis (MESA) dataset\textcolor{blue}{\hyperlink{ref18}{\cite{ref18}}} was used for model training, validation, and ID testing. The MESA sleep study comprises data from 2055 subjects (53.63\% female), aged between 54 and 95 years. Participants included in the study had not used treatments for sleep apnea, such as continuous positive airway pressure or oxygen devices, for more than a month prior to the study, or only used it less than once a week. 
 
The overnight PSG recordings obtained at home include PPG and SpO2 time series. The sampling rate of PPG signals is 256~Hz, whereas SpO2 is sampled at 1~Hz. Time labels and durations of apnea and hypopnea episodes in the respiratory flow signal were annotated by trained technicians. Due to potential PPG signal quality issues, we included only the 276~subjects whose PSG studies are labeled with the highest quality grade of~7, as in\textcolor{blue}{\hyperlink{ref16}{\cite{ref16}}}.

For the MESA dataset, the PSG study quality grades and corresponding AHI values were obtained from the \texttt{mesa-sleep-dataset-0.4.0.csv} file, specifically from the \texttt{overall5} and \texttt{ahi-a0h3a} columns, respectively. While the AB labels for each subject were computed based on the annotated apnea/hypopnea event onset time and durations provided in the corresponding XML files.

The MESA dataset has already been used in the following recent studies\textcolor{blue}{\hyperlink{ref16}{\cite{ref16}}},\textcolor{blue}{\hyperlink{ref19}{\cite{ref19}}},\textcolor{blue}{\hyperlink{ref19b}{\cite{ref19b}}},\textcolor{blue}{\hyperlink{ref19c}{\cite{ref19c}}}.

\subsubsection{OSASUD data}	 

To assess OOD generalization, we tested models on the open‑access Obstructive Sleep Apnea Stroke Unit Dataset (OSASUD)\textcolor{blue}{\hyperlink{ref20}{\cite{ref20}}}. This PSG dataset involves 30 after-stroke patients (36.67\% female) admitted due to suspected cerebrovascular events, such as ischemic and hemorrhagic stroke, or transient ischemic attack. Patients aged $<$18 years, those unable to comply with standard requirements for PSG monitoring, individuals with severe aphasia impacting understanding or consent, and those at high risk of alcohol or drug withdrawal syndrome were not included. Notably, conditions such as obesity, diabetes mellitus, atrial fibrillation and other cardiac diseases did not serve as exclusion criteria.

The dataset includes PSG signals such as finger PPG and SpO2. The sampling rate of PPG signals is 80~Hz, whereas SpO2 time series are sampled at 1~Hz. In addition, SpO2 values $<$50\% or $>$100\% were identified as artifacts and marked as null. 

The OSASUD data were annotated by a trained sleep medicine physician based on the sleep scoring guidelines established by the American Academy of Sleep Medicine, identifying occurrences of central, obstructive, mixed apnea, and hypopnea events (1~s~granularity). 

For the OSASUD dataset, the AHI labels were obtained from the original study\textcolor{blue}{\hyperlink{ref20}{\cite{ref20}}}. Meanwhile, the AB values were computed based on the annotated apnea and hypopnea events with a temporal resolution of 1~s.

The OSASUD dataset has already been used in the following recent studies\textcolor{blue}{\hyperlink{ref14c}{\cite{ref14c}}}\textcolor{blue}{\hyperlink{ref21}{\cite{ref21}}},\textcolor{blue}{\hyperlink{ref22}{\cite{ref22}}},\textcolor{blue}{\hyperlink{ref23}{\cite{ref23}}}. A comparison of the two datasets used is provided in Table~\textcolor{blue}{\ref{Table1}}.

\begin{table*}[t!]
	\centering
	\caption{Differences between MESA and OSASUD datasets. PSG--polysomnography,  ECG--electrocardiography, EEG--electroencephalography, EOG--electrooculography, EMG--electromyography.}
	\setlength{\tabcolsep}{8pt} 
	\renewcommand{\arraystretch}{1.2} 
	
	\begin{tabular}{p{4.5cm} p{5.5cm} p{5.5cm}}
		\hline 
		\textbf{Dataset differences} &
		\textbf{\#1 MESA} &
		\textbf{\#2 OSASUD} \\
		\hline
		
		\textbf{Number of subjects (female \%)} &
		2055 (53.63\%) &
		30 (36.67\%) \\
		
		\textbf{Subject age} &
		69.37 $\pm$ 9.12 years&
		68.97 $\pm$ 11.22 years\\
		
	    \textbf{Collecting time} &
		2010--2012&
		2019--2020\\
		
		\textbf{PSG signals} &
		PPG, ECG, EEG, EOG, EMG, SpO2, heart rate, nasal airflow, snoring, 3-axis accelerometer, abdominal and thoracic movements &
		PPG, ECG, SpO2, respiratory rate, perfusion index, heart rate, nasal airflow, snoring, 3-axis accelerometer, abdominal and thoracic movements \\
		
		\textbf{Ethnicity} &
		Diverse population: White, African American, Hispanic, and Chinese American participants &
		Not considered \\
		
		\textbf{Clinical status} &
		General population &
		After-stroke patients \\
		
		\textbf{Environmental conditions} &
		Home-based~PSG: Participants from 6 clinical centres &
		Hospital-based~PSG: Clinical Neurology Unit of Udine University Hospital, Italy \\
		
		\textbf{Device} &
		Compumedics Somte monitoring system (Compumedics Ltd., Australia) &
		Embla RemLogic (Natus Medical Inc., Pleasanton, USA) \\
		
		\textbf{PPG sensor} &
		Nonin 8000 sensor (finger PPG, sampling rate -- 256~Hz) &
		Embletta polysomnograph + Mindray monitoring system (finger PPG, sampling rate -- 80~Hz) \\
		
		\hline 
	\end{tabular}
	
	\label{Table1}
\end{table*}

\subsection{Dataset partitioning $\&$ preprocessing}

This study included a total of 306 subjects, consisting of 276 individuals from the MESA cohort and 30 patients from the OSASUD cohort. The MESA dataset was used for model training and ID evaluation, and partitioned into training (216 subjects), validation (30 subjects), and ID testing (30 subjects) subsets. The OSASUD dataset, containing 30 subjects from a clinically distinct population, was utilized for OOD evaluation to assess model robustness under domain shift. 

\begin{figure}[h] 
	\centering 
	\includegraphics[width=\columnwidth]{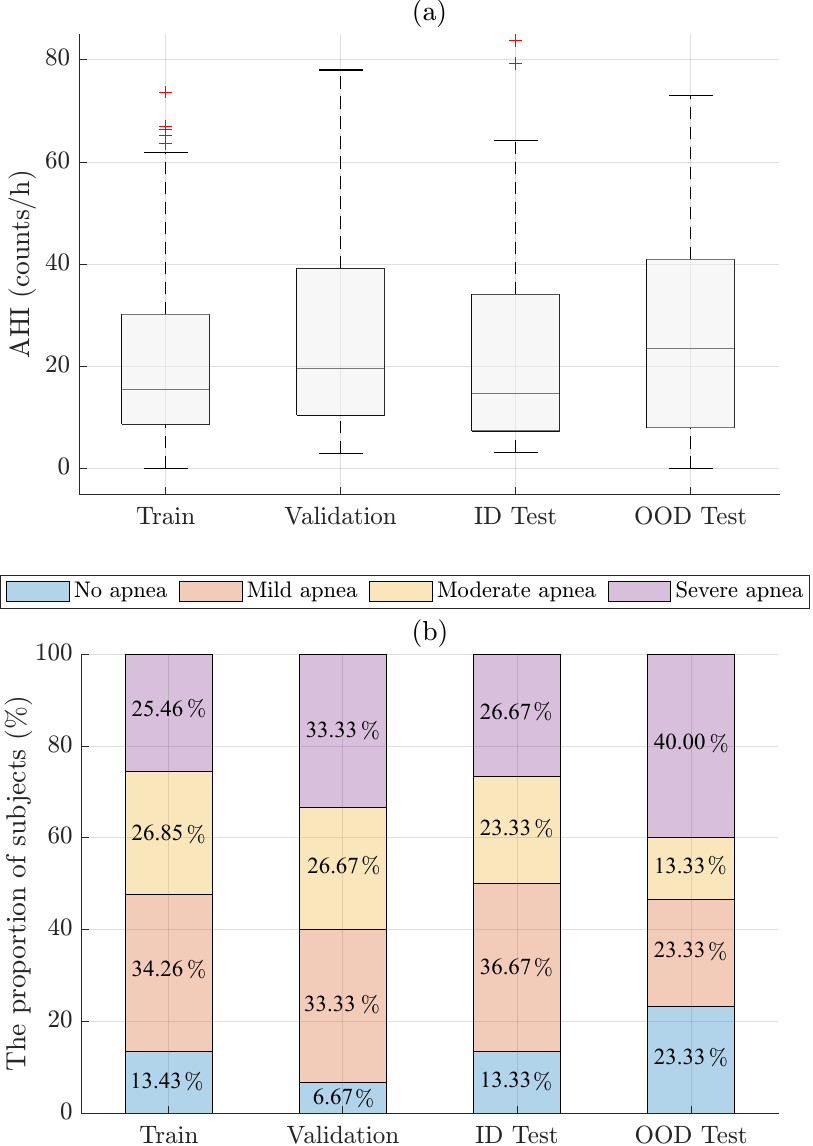}
	\caption{The distributions of AHI across dataset splits: (a) boxplots of AHI values for training, validation, ID test, and OOD test sets; (b) class distributions of subjects according to sleep apnea severity levels (no apnea, mild, moderate, and severe).}
	\label{Fig3}
\end{figure}

The distributions of AHI values in prepared training, validation, ID testing, and OOD testing data are provided in Figure~\textcolor{blue}{\ref{Fig3}}. As shown in Figure~\textcolor{blue}{\ref{Fig3}b}, the ID testing data is dominated by mild apnea cases, followed by severe, moderate, and no apnea cases. In contrast, the OOD testing data exhibits a higher proportion of severe apnea cases, whereas moderate apnea represents the smallest group.

In this study, the PPG signals were preprocessed using a zero-phase, fourth-order Butterworth band-pass filter (0.4--9~Hz) to mitigate noise and enhance fiducial point detection. These PPG signals were then resampled to $f_{\mathrm{s}}$~=~32~Hz using a low-pass finite impulse response filter to prevent aliasing. This sampling rate is sufficient to capture relevant cardiovascular dynamics\textcolor{blue}{\hyperlink{ref23b}{\cite{ref23b}}} and PPG morphology changes while reducing computational load. In addition, our experiments have shown that the PPG sampling frequency, provided that the Nyquist criterion ($f_{\max}$~$\leq$~$f_{\mathrm{s}}$/2, where $f_{\max}$~$\approx$~9--10~Hz) is satisfied, does not have a significant impact on the performance of PPG feature-based models. To mitigate amplitude-related distributional differences between datasets, overnight PPG recordings from each subject were independently standardized using root-mean-square (RMS) normalization prior to feature extraction. 

\subsection{Model predictors $\&$ segmentation}

\begin{figure}[h] 
	\centering 
	\includegraphics[width=\columnwidth]{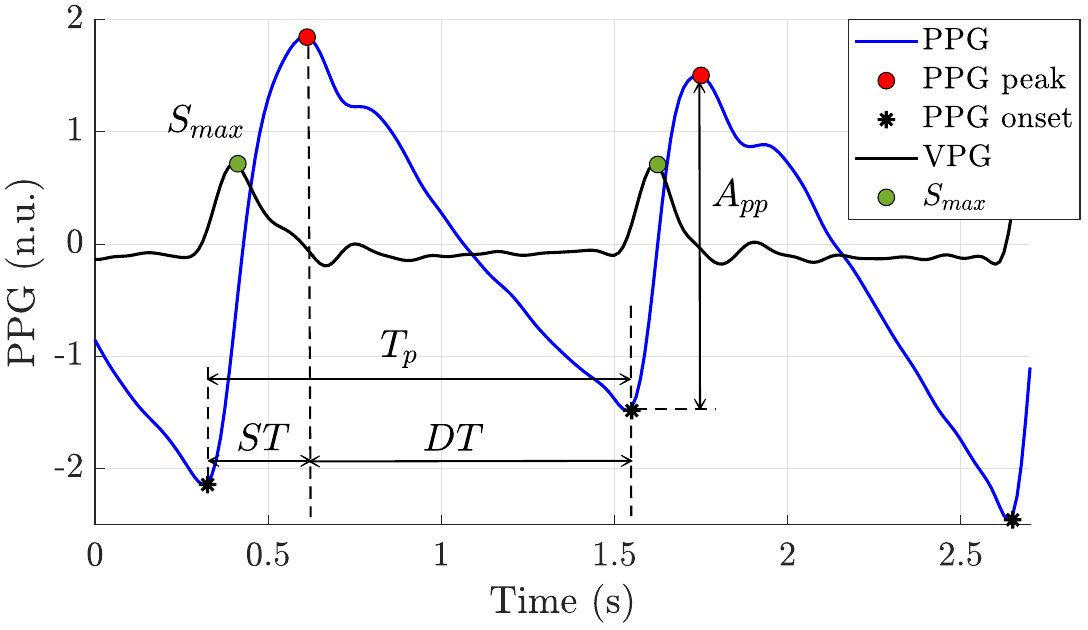} 
	\caption{PPG pulse wave feature delineation: pulse wave interval, $T_p$, systolic time, $ST$, diastolic time, $DT$, peak-to-peak amplitude, $A_{pp}$, and maximum slope of PPG, $S_{max}$, calculated from the first derivative of PPG signal, VPG (velocity photoplethysmogram).}
	\label{Fig4}
\end{figure}

\begin{figure}[h] 
	\centering 
	\includegraphics[width=\columnwidth]{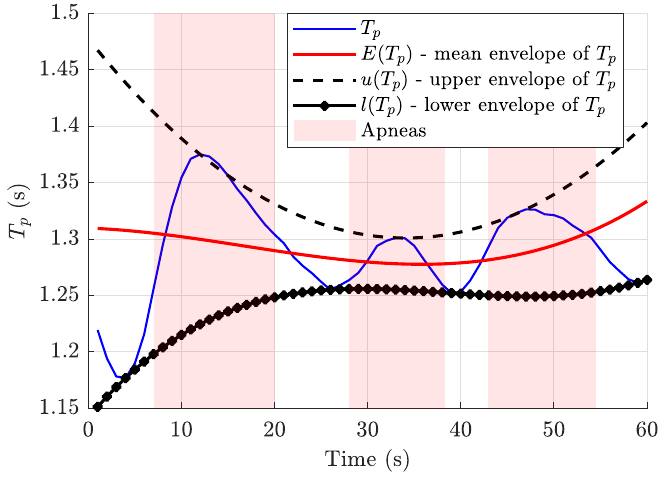} 
	\caption{The example of the proposed PPG feature, the mean envelope of pulse wave interval time series, $E$($T_p$), with marked apnea episodes.}
	\label{Fig5}
\end{figure}

To calculate the PPG-derived features, PPG peaks and onset points were identified using an automatic beat detection algorithm proposed in\textcolor{blue}{\hyperlink{ref24}{\cite{ref24}}}. We extracted five pulse wave features (see Figure~\textcolor{blue}{\ref{Fig4}}) and one proposed higher-level feature (see Figure~\textcolor{blue}{\ref{Fig5}}):

\begin{enumerate}
	\item pulse wave interval, $T_p$;
	\item peak-to-peak amplitude, $A_{pp}$;
	\item systolic time, $ST$;
	\item diastolic time, $DT$;
	\item maximum slope of PPG, $S_{max}$;
	\item mean envelope of pulse wave interval, $E(T_p)$.
\end{enumerate}

\vspace{2pt}
The listed features were inspired by previous studies\textcolor{blue}{\hyperlink{ref6}{\cite{ref6}}},\textcolor{blue}{\hyperlink{ref7}{\cite{ref7}}},\textcolor{blue}{\hyperlink{ref16}{\cite{ref16}}}. To evaluate the contribution of PPG-derived features, three different input combinations were investigated: (i) SpO2 combined with all six PPG features, including the proposed mean envelope feature $E(T_p)$, (ii) SpO2 combined with five PPG pulse wave features ($T_p$, $ST$, $DT$, $A_{pp}$, $S_{max}$), and (iii) SpO2 only.

\begin{figure*}[!t] 
	\centering 
	\includegraphics[width=14.5cm]{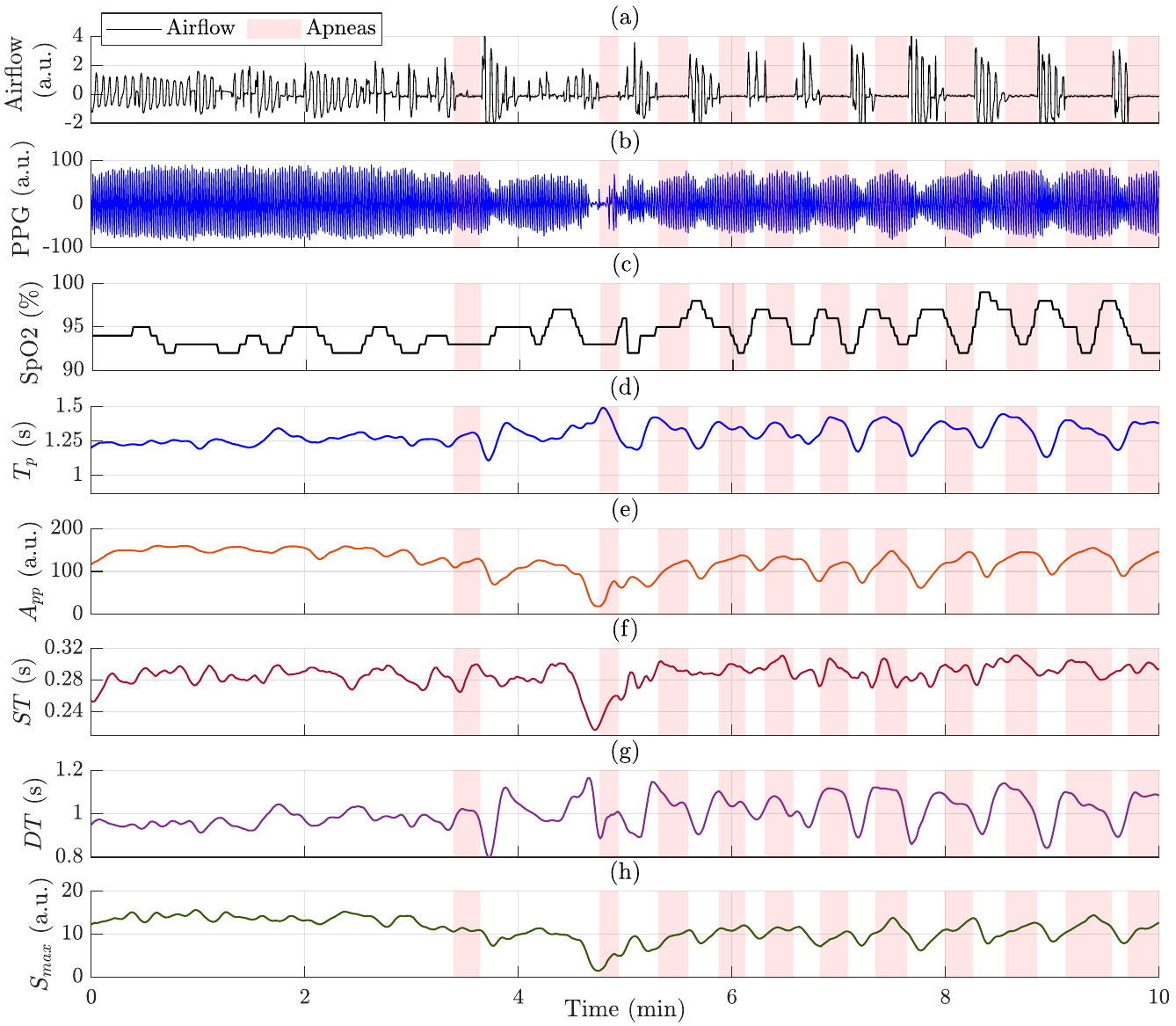} 
	\caption{The changes of PPG feature sequences and SpO2 during apnea events: (a) respiratory airflow with marked apnea episodes, (b) PPG signal, (c) SpO2, (d) pulse wave interval, $T_p$, (e) peak-to-peak amplitude, $A_{pp}$, (f) systolic time, $ST$, (g) diastolic time, $DT$, and (h) maximum slope of PPG, $S_{max}$.}
	\label{Fig6}
\end{figure*}

These feature sequences were resampled to 1~Hz and then smoothed with a fifth-order moving-average filter. The typical variability of PPG feature sequences and SpO2 during sleep apnea episodes is illustrated in Figure~\textcolor{blue}{\ref{Fig6}}.

Furthermore, we assume that the maximum slope of PPG, $S_{max}$, defined as the maximum value of its first derivative, may provide complementary information for evaluating respiratory disturbances. This feature reflects the maximum rate of blood volume changes during the systolic upstroke and may capture hemodynamic variations influenced by autonomic nervous system activity associated with sleep apnea. To the best of our knowledge, $S_{max}$ has not been previously explored in this context, and it is therefore investigated in this study.

This study also proposes a novel PPG feature, the mean envelope of the pulse wave interval time series, $E$($T_p$), defined as the average of the upper and lower envelopes of the sequence within a 1-minute epoch (see Figure~\textcolor{blue}{\ref{Fig5}}). The upper $u(T_p)$ and lower $l(T_p)$ envelopes are obtained by identifying local maxima and minima of the $T_p$ sequence and interpolating between these extrema. Specifically, a peak-based method is used, where the envelopes are constructed from the detected peaks and troughs using a shape-preserving spline interpolation scheme.

The mean envelope is then computed as:
\begin{equation}
	E(T_p) = \frac{u(T_p) + l(T_p)}{2}.
\end{equation}

The mean envelope of the PPG pulse-to-pulse interval, $E$($T_p$), reflects low-frequency, slowly varying autonomic modulation of heart rate. Because pulse-to-pulse intervals shorten after apnea‑induced sympathetic activation with a characteristic delay during arousal and lengthen during recovery (see Figure~\textcolor{blue}{\ref{Fig6}d}), this proposed feature captures the minute-scale cardiovascular response to sleep apnea events and overall autonomic balance.

The analyzed six PPG features and SpO2 were divided into 5-minute segments. Each feature was then separately standardized according to the mean and standard deviation of the training data. We opted for a trade-off 5-minute segmentation because 1-minute and smaller windows are likely too short to capture meaningful associations among PPG-derived features and SpO2 changes during sleep apnea events. While using much longer segments increases the computational load during model training.

\subsection{Model architectures}
\label{Models}

Three artificial neural network architectures were investigated to evaluate different strategies for capturing temporal dependencies in PPG-derived features and SpO2 signals for AB prediction, and to assess the influence of model complexity on performance and generalization. The selected architectures represent different modeling paradigms: hybrid convolutional-recurrent models, namely 2CNN-GRU and 2CNN-2LSTM, with two sequential convolutional layers, followed by gated recurrent unit (GRU) and two stacked long short-term memory (LSTM) layers, respectively, and a purely convolutional multi-scale model (parallel CNNs).

\subsubsection{2CNN-GRU architecture}

\begin{figure*}[!t]
	\centering 
	\includegraphics[width=\textwidth]{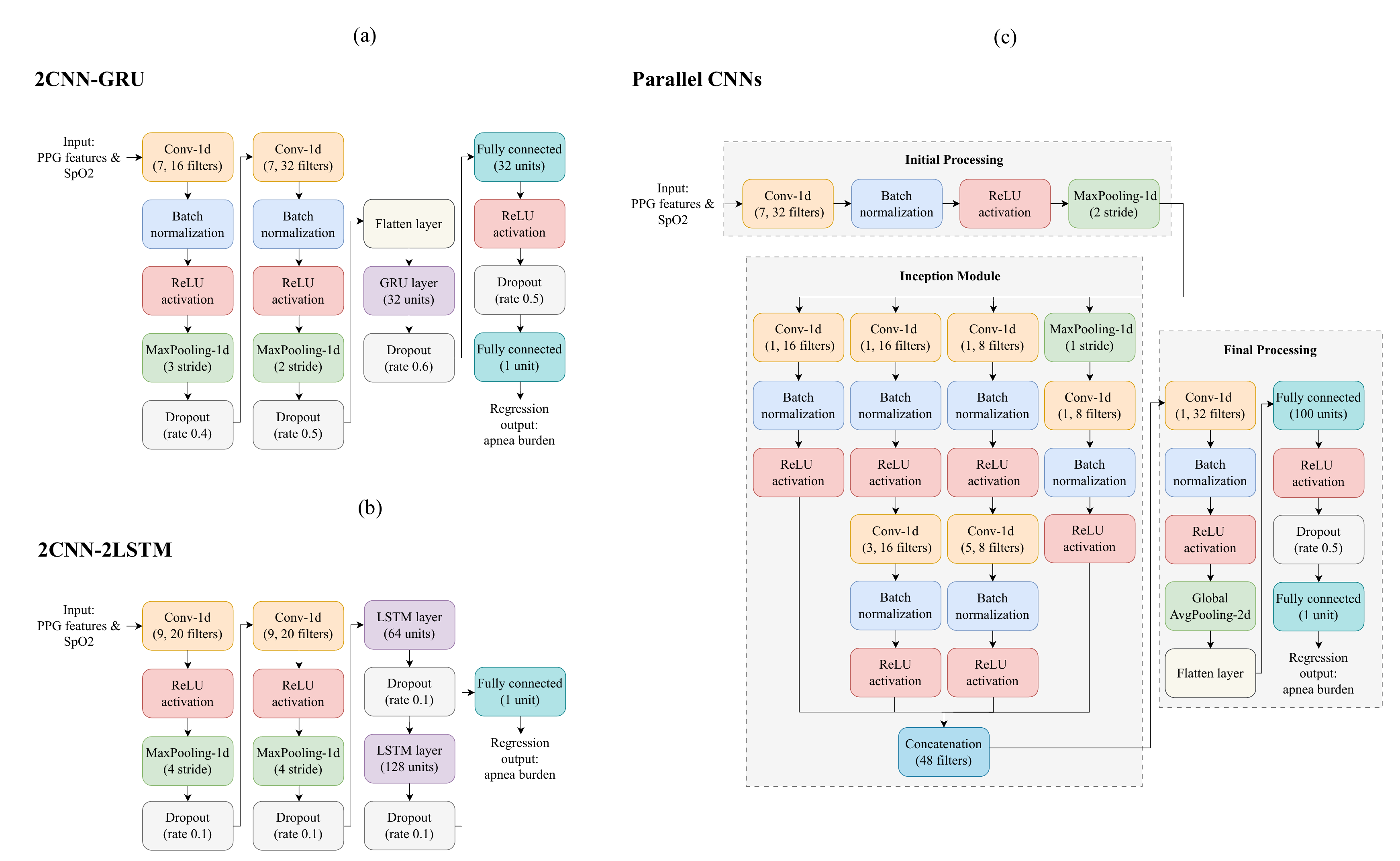} 
	\caption{The block diagrams of the implemented model architectures: (a) 2CNN-GRU, (b) 2CNN-2LSTM, and (c) parallel CNNs.}
	\label{Fig7}
\end{figure*}

The proposed 2CNN-GRU architecture combines convolutional feature extraction with gated recurrent temporal modeling (see Figure~\textcolor{blue}{\ref{Fig7}a}). Convolutional layers capture local patterns within short segments of the input features, whereas the GRU layer models temporal dependencies by learning how these patterns evolve over time in a computationally efficient manner. Compared to LSTM-based models, GRU offers reduced complexity and faster training, which is advantageous for home-based monitoring applications. This architecture was designed to provide a balance between performance and model simplicity.

\subsubsection{2CNN-2LSTM architecture}

The 2CNN-2LSTM architecture follows a commonly used hybrid approach that integrates convolutional feature extraction with stacked LSTM layers for modeling long-range temporal dependencies (see Figure~\textcolor{blue}{\ref{Fig7}b}). This type of architecture has been widely applied in physiological signal analysis and is inspired by previously proposed frameworks for PPG-based blood pressure and heart rate estimation\textcolor{blue}{\hyperlink{ref25}{\cite{ref25}}}. It serves as a reference model to evaluate the impact of capturing more complex sequential dependencies compared to the proposed GRU-based architecture.

\subsubsection{Parallel CNNs architecture}

The parallel CNNs architecture is an Inception-inspired one-dimensional convolutional model (see Figure~\textcolor{blue}{\ref{Fig7}c}) designed to capture multi-scale temporal features through parallel processing paths\textcolor{blue}{\hyperlink{ref26}{\cite{ref26}}}. By extracting features at multiple temporal resolutions simultaneously, this architecture is particularly suitable for physiological signals, where relevant patterns may occur at different time scales. This model was included to assess whether convolutional multi-scale representations can compete with hybrid convolutional-recurrent approaches.

\subsubsection{Parameter settings}

The architectural hyperparameters were selected empirically based on preliminary experiments, with the final configuration chosen according to validation set performance.

The implemented models were trained using identical optimization settings, which were chosen after some minor tuning. The selected model training parameters are provided in Table~\textcolor{blue}{\ref{TableTraining}}. The models were trained using the mean squared error (MSE) loss function to predict continuous AB values.

\begin{table}[h!]
	\centering
	\caption{Model training settings.}
	\renewcommand{\arraystretch}{1.2} 
	\setlength{\tabcolsep}{6pt}
	
	\begin{tabular}{l l}
		\hline
		\textbf{Parameter} & \textbf{Value} \\
		\hline
		
		Optimizer & Adam \\
		Learning rate & 0.001 \\
		Mini-batch size & 1024 \\
		Maximum epochs & 1000 \\
		Early stopping (patience) & 15 epochs \\
		Loss function & Mean squared error (MSE) \\
		
		\hline
	\end{tabular}
	
	\label{TableTraining}
\end{table}

The mini-batch size was selected based on performed experiments, which indicated that this parameter had a negligible effect on the final model performance. Therefore, a larger mini-batch size of 1024 was chosen to reduce computational time and improve training efficiency. 

The learning rate was empirically optimized by comparing values of 0.1, 0.01, 0.001, and 0.0001. Among these, a learning rate of 0.001 consistently provided the best performance across implemented models and was therefore used in our experiments.

\subsection{Severity assessment $\&$ Performance evaluation}
\label{SeverityAssessment}
To define the labels, the AB value was calculated for each 5-minute segment and then for the entire signal for each subject separately. In our study, the segment-level AB is defined as:

\begin{equation}
	\mathrm{AB}_{\mathrm{segment}} = \frac{\sum d_i}{T_{\mathrm{seg}}},
\end{equation}

where $\sum d_i$ denotes the total duration of apnea and hypopnea events within the segment, and $T_{\mathrm{seg}}$ is the segment length (in seconds).

The subject-level AB is defined by averaging the segment-wise values:

\begin{equation}
	\mathrm{AB} = \frac{1}{N} \sum_{k=1}^{N} \mathrm{AB}_{\mathrm{segment}}^{k},
\end{equation}
\label{AB_eq}

where $N$ is the total number of segments in the recording.

The implemented models first predict AB for each 5‑minute segment based on the extracted features. These predictions are then aggregated over the entire recording for each subject to estimate the subject‑level AB (see Equation~\textcolor{blue}{\ref{AB_eq}}), which is then converted into the AHI for severity assessment. The predicted AHI is computed as:

\begin{equation}
	\mathrm{AHI}_{\mathrm{AB}} = \frac{{\mathrm{AB}}\cdot 3600}{D_{\mathrm{mean}}},
\end{equation}
\label{AHI_AB}

where $\mathrm{AB}$ denotes the overall predicted AB, and $D_{\mathrm{mean}}$ represents the mean duration of apnea/hypopnea events (in seconds) calculated from overnight recordings of the training data.

\begin{figure}[h] 
	\centering 
	\includegraphics[width=\columnwidth]{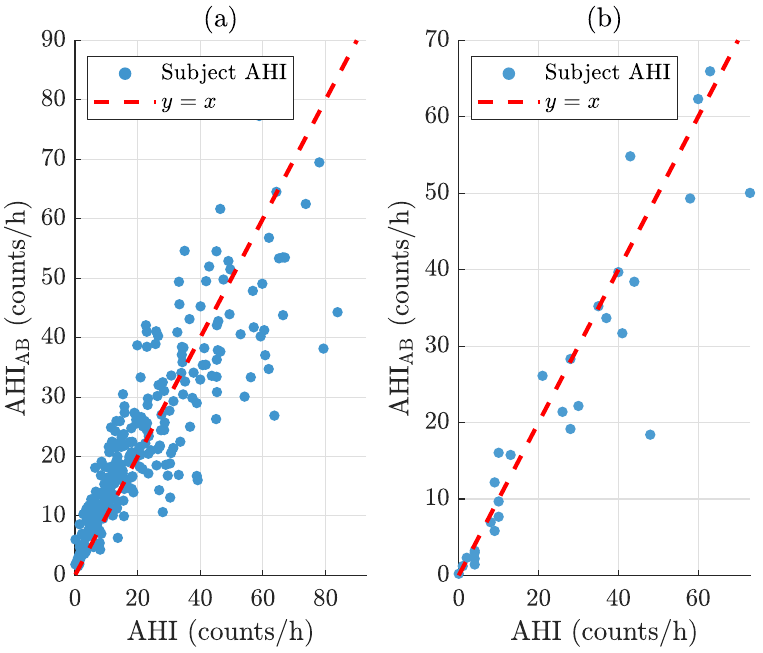} 
	\caption{Scatter plots of reference AHI versus AB-derived AHI ($\mathrm{AHI}_{\mathrm{AB}}$): (a) MESA data (276 subjects); (b) OSASUD data (30 subjects). The dashed red line ($y = x$) indicates ideal agreement between the two measures.}
	\label{Fig2}
\end{figure}

This conversion is justified by the approximately linear relationship between AB and AHI observed in the datasets, as shown in Figure~\textcolor{blue}{\ref{Fig2}}. Specifically, the scatter plots show that $\mathrm{AHI}_{\mathrm{AB}}$ closely follows the reference AHI values along the identity line ($y$~=~$x$), indicating that AB provides a consistent surrogate for event frequency when normalized by the average event duration.

The mean absolute error with standard deviation between the estimated $\mathrm{AHI}_{\mathrm{AB}}$ and the ground-truth AHI is 6.79~$\pm$~6.67 counts/h for MESA data, and 4.99~$\pm$~6.66 counts/h for OSASUD data, indicating the error introduced by the fixed-duration AB-to-AHI conversion.

Based on the estimated $\mathrm{AHI}_{\mathrm{AB}}$, the subjects were categorized into four sleep apnea severity groups (see Subsection~\textcolor{blue}{\ref{Dataset}}), enabling a multi-classification framework. This evaluation was performed for both ID and OOD test-subsets.

To assess model prediction performance, regression metrics were evaluated. These include mean absolute error (MAE) and mean absolute scaled error (MASE) for both predicted AB and estimated AHI.

The MAE is defined as:
\begin{equation}
	\mathrm{MAE} = \frac{1}{N} \sum_{i=1}^{N} |y_i - \hat{y}_i|,
\end{equation}
where $y_i$ is the ground-truth value, $\hat{y}_i$ is the predicted value, and $N$ is the number of samples.

The MASE was computed as:
\begin{equation}
	\mathrm{MASE} = \frac{\mathrm{MAE}}{\mathrm{MAE}_{\mathrm{baseline}}},
\end{equation}
where $\mathrm{MAE}_{\mathrm{baseline}}$ denotes the MAE obtained by a constant predictor equal to the median AB or AHI value from the testing data.

To assess multi-classification performance\textcolor{blue}{\hyperlink{ref26b}{\cite{ref26b}}}, macro-averaged metrics were computed, including sensitivity, specificity, positive predictive value (PPV), accuracy, and F1-score. Macro-averaging computes each metric independently per class and then averages them, assigning equal importance to all classes. Additionally, micro-accuracy was calculated as the total number of correct predictions divided by the total number of samples, such that classes with a larger number of samples contribute more to the overall score.

In addition, agreement metrics including Matthews correlation coefficient (MCC) and Cohen’s kappa were evaluated. The extended MCC for multi-classification was computed using the confusion matrix formulation proposed by Gorodkin\textcolor{blue}{\hyperlink{ref27}{\cite{ref27}}}, \textcolor{blue}{\hyperlink{ref28}{\cite{ref28}}}:
\begin{equation}
	\mathrm{MCC} = \frac{\mathrm{trace}(\mathbf{C}) \cdot S - \sum_{k} r_k c_k}{\sqrt{\left(S^2 - \sum_{k} c_k^2\right)\left(S^2 - \sum_{k} r_k^2\right)}},
\end{equation}
where $\mathbf{C}$ is the confusion matrix, $S = \sum_{i,j} C_{ij}$ is the total number of samples, $r_k$ and $c_k$ denote the row and column sums corresponding to class $k$, representing the total number of true and predicted samples for that class, respectively, and $\mathrm{trace}(\mathbf{C})$ is the sum of diagonal elements, i.e., the total number of correct predictions across all classes.

Cohen’s kappa coefficient, originally introduced in\textcolor{blue}{\hyperlink{ref29}{\cite{ref29}}}, can be used to measure agreement between predicted and true labels while accounting for chance agreement. Cohen’s kappa is defined as:
\begin{equation}
	\kappa = \frac{p_o - p_e}{1 - p_e},
\end{equation}
where $p_o = \frac{\mathrm{trace}(\mathbf{C})}{N}$ is the observed agreement, and 
\begin{equation}
	p_e = \frac{\sum_{k} r_k c_k}{N^2}
\end{equation}
is the expected agreement by chance, with $N$ being the total number of samples.

All performance metrics were computed over five independent training runs with different fixed random seeds to ensure reproducibility. The final results are reported as mean~$\pm$~standard deviation, providing an estimate of model stability.

\section{Results}
\label{Section3}
\subsection{Data-extracted statistics $\&$ baseline errors}

From the training subset, the mean duration of apnea/hypopnea events was calculated as $D_{\mathrm{mean}}$~=~21.39~s. This value was subsequently used to convert predicted AB into the AHI metric. Additionally, the median values of the ID testing set were $\mathrm{median}(\mathrm{AB}_\mathrm{ID})$~=~11.22\% and $\mathrm{median}(\mathrm{AHI}_\mathrm{ID})$~=~14.70~counts/h. While the median values of the OOD testing set were $\mathrm{median}(\mathrm{AB}_\mathrm{OOD})$~=~11.16\% and $\mathrm{median}(\mathrm{AHI}_\mathrm{OOD})$~=~23.50~counts/h. These median values were used as baseline predictors for calculating the MASE.

Table~\textcolor{blue}{\ref{Table2}} summarizes the baseline mean absolute errors for AB and AHI estimation obtained when always predicting the median values from the ID and OOD testing data, respectively. These baseline errors were used for normalization in MASE calculation.

\begin{table}[h!]
	\centering
	\caption{Mean absolute errors (baseline) for ID and OOD testing cases.}
	\setlength{\tabcolsep}{6pt}
	\renewcommand{\arraystretch}{1.2}
	
	\begin{tabular}{lcc}
		\hline
		\textbf{Mean absolute errors (baseline)} & \textbf{ID testing} & \textbf{OOD testing} \\
		\hline
		MAE(AB) (\%) & 8.05 & 9.35 \\
		MAE(AHI) (counts/h) & 15.50 & 18.17 \\
		\hline
	\end{tabular}
	\label{Table2}
\end{table}

\subsection{Model prediction errors}

\begin{figure*}[!t]  
	\centering 
	\includegraphics[width=15.5cm]{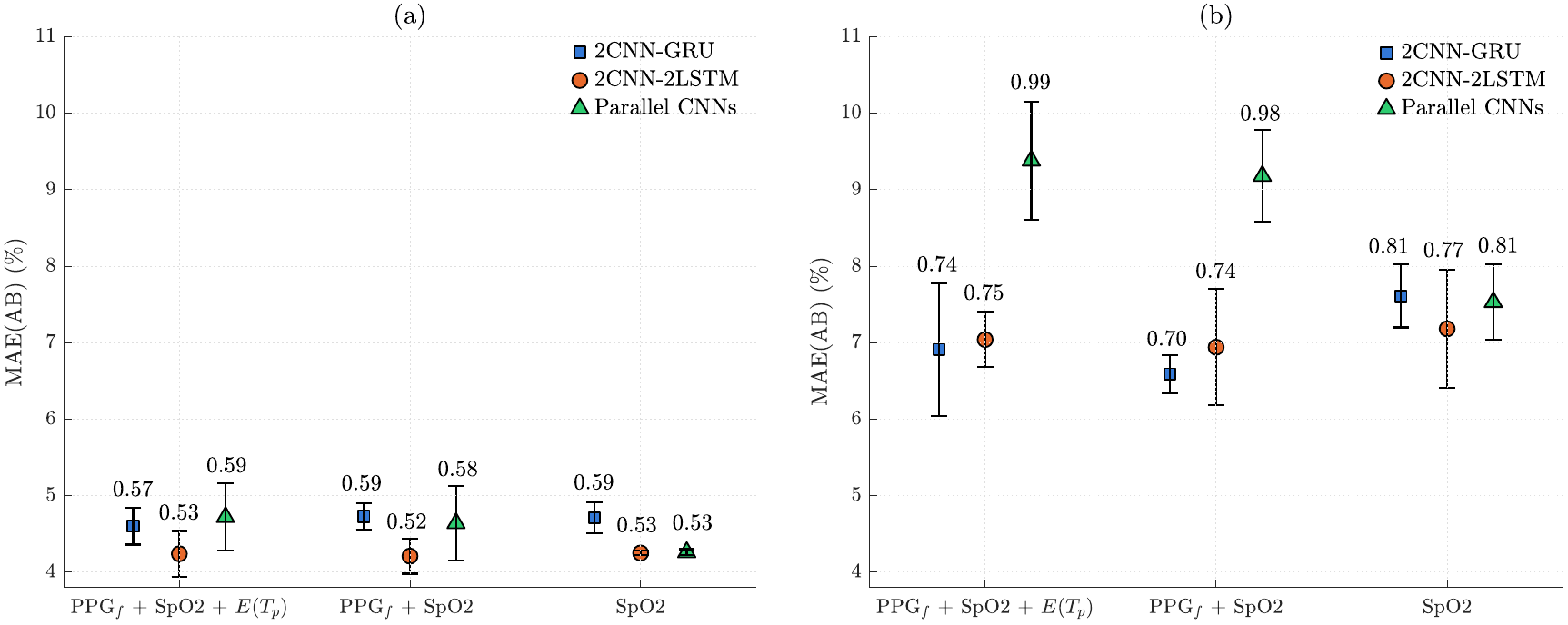} 
	\caption{Mean absolute errors of predicted AB when using different model architectures and predictor combinations for (a) ID testing and (b) OOD testing data. Error bars (mean $\pm$ standard deviation) obtained from five iterations when different fixed random seeds were used. PPG$_f$ corresponds to five pulse wave features, whereas $E(T_p)$ -- the mean envelope feature of the pulse wave interval. Mean absolute scaled errors are shown above error bars.}
	\label{Fig17}
\end{figure*}

\begin{figure*}[!t]
	\centering 
	\includegraphics[width=15.5cm]{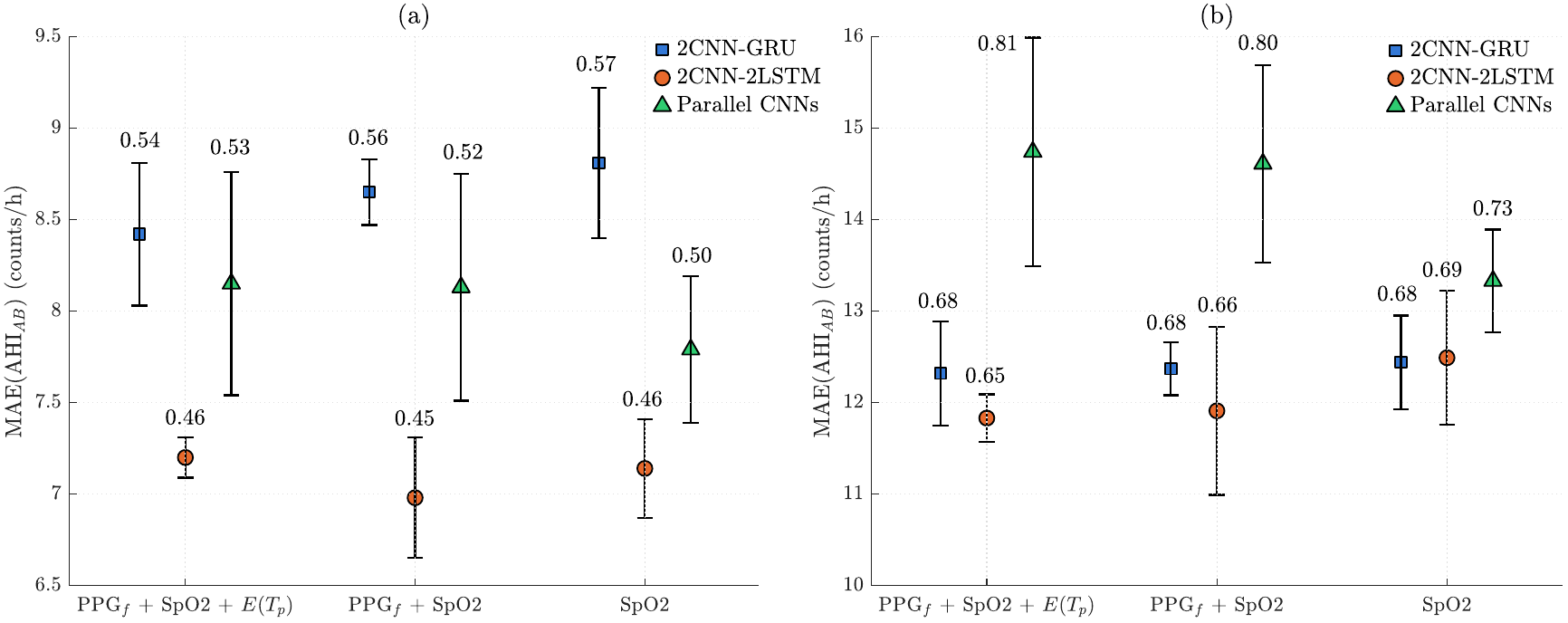} 
	\caption{Mean absolute errors of predicted AHI when using different model architectures and predictor combinations for (a) ID testing and (b) OOD testing data. Error bars (mean $\pm$ standard deviation) obtained from five iterations when different fixed random seeds were used. PPG$_f$ corresponds to five pulse wave features, whereas $E(T_p)$ -- the mean envelope feature of the pulse wave interval. Mean absolute scaled errors are shown above error bars.}
	\label{Fig18}
\end{figure*}

In terms of regression performance, the lowest MAE(AB) in ID testing was achieved by the 2CNN-2LSTM model regardless of the features used (see Figure~\textcolor{blue}{\ref{Fig17}}), whereas the lowest OOD MAE(AB) was obtained by the 2CNN-GRU model using PPG pulse wave features and SpO2, with a value of 6.59~$\pm$~0.25\% (MASE~=~0.70). For AHI estimation (see Figure~\textcolor{blue}{\ref{Fig18}}), the lowest ID MAE($\mathrm{AHI}_{\mathrm{AB}}$) and OOD MAE($\mathrm{AHI}_{\mathrm{AB}}$) were also achieved by the 2CNN-2LSTM model. In contrast, the parallel CNNs architecture produced the largest OOD regression errors, particularly when PPG features were included. Regarding the comparison of estimating AB and AHI (see Figures~\textcolor{blue}{\ref{Fig17}}~and~\textcolor{blue}{\ref{Fig18}}), it can be observed that in all cases $\mathrm{MASE}(\mathrm{AHI}_{\mathrm{AB}}) < \mathrm{MASE}(\mathrm{AB})$, indicating that the relative error of AHI estimation is lower than that of AB.

\subsection{Subject-level classification performance}

\begin{figure*}[!t]
	\centering 
	\includegraphics[width=\textwidth]{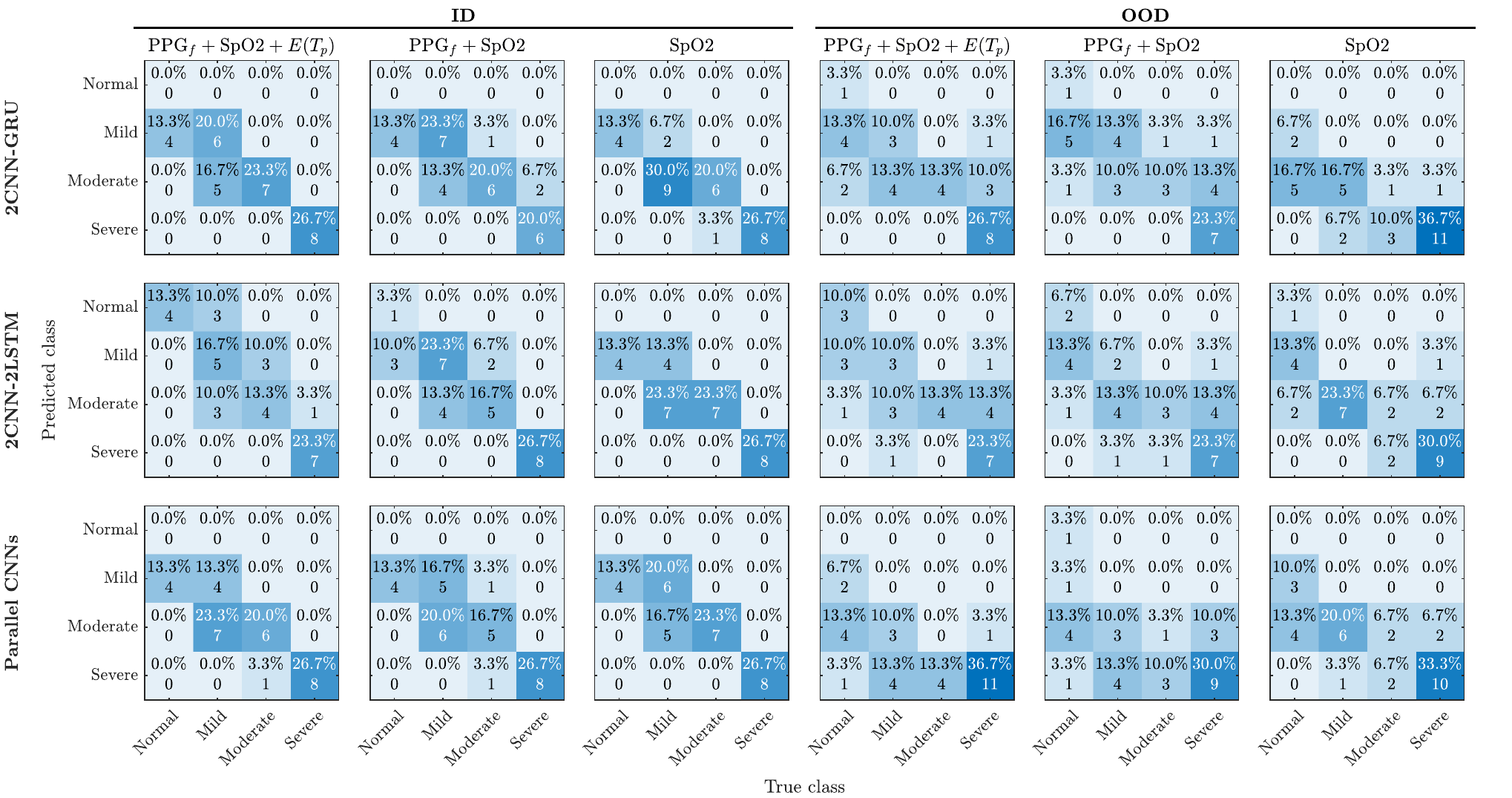} 
	\caption{Confusion matrices for classifying sleep apnea patients into four severity groups (normal, mild, moderate, and severe) using three model architectures and different predictor combinations. The left panel shows ID, whereas the right panel presents OOD testing results. Each matrix reports class-wise prediction percentages (top) and the corresponding number of subjects (bottom). Predictor combinations include five PPG pulse wave features (PPG$_f$) with SpO2 and the proposed mean envelope feature $E(T_p)$, PPG$_f$ with SpO2, and SpO2 alone.}
	\label{FigCM}
\end{figure*}

As illustrated in Figure~\textcolor{blue}{\ref{FigCM}}, the confusion matrices provide a detailed view of class-wise prediction behavior across different model architectures and predictor combinations for both ID and OOD testing scenarios. In general, the ID results exhibited a more pronounced diagonal structure, indicating more accurate classification, whereas the OOD results showed increased dispersion of predictions, particularly between neighboring severity classes. Misclassifications are most frequently observed between mild and moderate groups, reflecting the inherent difficulty of distinguishing adjacent apnea severity levels. Additionally, the inclusion of PPG-derived features led to more concentrated diagonal patterns compared to using SpO2 alone, suggesting improved class separability.

The subject-level classification performance obtained using different model architectures and predictor combinations is summarized in Table~\textcolor{blue}{\ref{Table_Metrics}}. Among the evaluated architectures, the 2CNN-2LSTM model achieved the highest overall performance, particularly in terms of macro-level sensitivity, PPV, and F1-score metrics. This model consistently outperformed the other architectures across both ID and OOD testing scenarios, indicating its stronger capability to capture long-range temporal dependencies in PPG-derived features and SpO2. Although this highest performance may also be partially influenced by hyperparameter tuning. For instance, it reached the highest macro-accuracy values of 76.52\% and 65.36\%, for ID and OOD respectively, when combining PPG pulse wave features, SpO2, and the proposed feature $E(T_p)$.

The 2CNN-GRU model also benefited from the inclusion of PPG-derived features and showed competitive performance overall. Notably, in OOD testing, the addition of PPG features improved macro-sensitivity from 30.00\% to 46.07\%, macro-accuracy from 55.15\% to 64.52\%, and macro-F1-score from 23.63\% to 36.70\% compared to using SpO2 alone. This highlights the importance of incorporating PPG-based information, particularly under distribution shifts.

In contrast, the parallel CNNs architecture exhibited different behavior. While it achieved competitive ID performance, particularly when only SpO2 was used, its overall performance was lower compared to hybrid architectures. Moreover, this model showed the weakest generalization to OOD data, with a noticeable degradation in sensitivity and F1-score. This suggests that purely convolutional approaches, even when incorporating multi-scale analysis, may be less effective in capturing relations between SpO2 and PPG features compared to hybrid convolutional-recurrent models. 

Across all models, combining PPG-derived features with SpO2 consistently led to improved OOD performance compared to using SpO2 alone. The proposed feature $E(T_p)$ provided an additional, albeit modest, benefit, especially for hybrid architectures such as the 2CNN-GRU. These findings demonstrate that PPG-derived features provide complementary physiological information that can be particularly valuable under distribution changes.

A general decrease in performance was observed when transitioning from ID to OOD testing across all models and feature combinations, highlighting the challenges of robust generalization to unseen datasets. Finally, it is worth noting that micro-accuracy values were consistently higher than macro-based metrics, reflecting the class imbalance present in the data and indicating that macro-level evaluation provides a more balanced assessment of model performance across severity groups.

\begin{table*}[t!]
	\centering
	\caption{Performance metrics obtained for classifying sleep apnea patients into four severity groups using different model architectures and predictor combinations. Averaged results (mean $\pm$ standard deviation) were obtained from five iterations with different fixed random seeds. PPG$_f$ corresponds to five pulse wave features, whereas $E(T_p)$ denotes the mean envelope feature of the pulse wave interval. The highest values for each metric in ID and OOD testing cases within each model are highlighted in bold.}
	\setlength{\tabcolsep}{7pt} 
	\renewcommand{\arraystretch}{1.5} 
	
	\begin{tabular}{llcccccc}
		\hline
		\multirow{2}{*}{\textbf{Model}} & \multirow{2}{*}{\textbf{Metric (\boldmath{$\%$})}}
		& \multicolumn{2}{c}{\boldmath{$PPG_f$} + \textbf{SpO2} + \boldmath{$E(T_p)$}} 
		& \multicolumn{2}{c}{\boldmath{$PPG_f$} + \textbf{SpO2}} 
		& \multicolumn{2}{c}{\textbf{SpO2}}  \\
		\cline{3-8}
		& & \textbf{ID} & \textbf{OOD} & \textbf{ID} & \textbf{OOD} & \textbf{ID} & \textbf{OOD} \\
		\hline
		
		\multirow{6}{*}{\textbf{2CNN-GRU}}
		& Macro-Sensitivity & \textbf{58.97 $\pm$ 4.03} & \textbf{46.07 $\pm$ 11.12} & 57.46 $\pm$ 1.74 & 45.00 $\pm$ 5.16 & 56.56 $\pm$ 4.03 & 30.00 $\pm$ 1.86 \\
		& Macro-Specificity & \textbf{87.51 $\pm$ 1.69} & \textbf{82.97 $\pm$ 2.54} & 86.99 $\pm$ 0.78 & 82.57 $\pm$ 1.26 & 86.03 $\pm$ 1.73 & 80.31 $\pm$ 0.63 \\
		& Macro-PPV         & \textbf{51.22 $\pm$ 4.47} & 40.64 $\pm$ 17.55 & 51.04 $\pm$ 2.26 & \textbf{42.66 $\pm$ 13.27} & 46.91 $\pm$ 4.66 & 21.04 $\pm$ 2.02 \\
		& Macro-Accuracy    & \textbf{73.24 $\pm$ 2.85} & \textbf{64.52 $\pm$ 6.81} & 72.23 $\pm$ 1.22 & 63.79 $\pm$ 3.10 & 71.29 $\pm$ 2.88 & 55.15 $\pm$ 1.13 \\
		& Macro-F1-score    & \textbf{53.29 $\pm$ 4.36} & \textbf{36.70 $\pm$ 9.89} & 52.46 $\pm$ 2.10 & 36.05 $\pm$ 5.83 & 49.12 $\pm$ 4.74 & 23.63 $\pm$ 1.29 \\
		& Micro-Accuracy    & \textbf{82.33 $\pm$ 2.53} & \textbf{72.67 $\pm$ 3.03} & 81.67 $\pm$ 1.18 & 71.33 $\pm$ 2.17 & 80.00 $\pm$ 2.64 & 69.33 $\pm$ 1.49 \\
		
		\hline
		
		\multirow{6}{*}{\textbf{2CNN-2LSTM}}
		& Macro-Sensitivity & \textbf{64.37 $\pm$ 6.78} & \textbf{47.62 $\pm$ 8.45} & \textbf{64.37 $\pm$ 3.76} & 46.07 $\pm$ 4.28 & 61.82 $\pm$ 2.96 & 31.90 $\pm$ 3.56 \\
		& Macro-Specificity & \textbf{88.67 $\pm$ 1.82} & \textbf{83.11 $\pm$ 2.21} & 88.62 $\pm$ 1.22 & 83.07 $\pm$ 1.19 & 88.43 $\pm$ 1.42 & 80.37 $\pm$ 0.98 \\
		& Macro-PPV         & 63.46 $\pm$ 9.03 & 55.93 $\pm$ 5.95 & \textbf{63.87 $\pm$ 11.50} & \textbf{57.38 $\pm$ 5.23} & 52.75 $\pm$ 2.97 & 38.08 $\pm$ 15.76 \\
		& Macro-Accuracy    & \textbf{76.52 $\pm$ 4.13} & \textbf{65.36 $\pm$ 5.32} & 76.50 $\pm$ 2.31 & 64.57 $\pm$ 2.72 & 75.13 $\pm$ 2.19 & 56.14 $\pm$ 2.24 \\
		& Macro-F1-score    & 61.70 $\pm$ 6.26 & \textbf{44.83 $\pm$ 6.94} & \textbf{62.35 $\pm$ 6.53} & 43.33 $\pm$ 3.70 & 55.49 $\pm$ 3.55 & 28.82 $\pm$ 6.18 \\
		& Micro-Accuracy    & \textbf{84.00 $\pm$ 2.53} & \textbf{74.00 $\pm$ 3.03} & \textbf{84.00 $\pm$ 1.90} & 73.67 $\pm$ 1.39 & 83.67 $\pm$ 2.17 & 69.00 $\pm$ 1.49 \\
		
		\hline
		
		\multirow{6}{*}{\textbf{Parallel CNNs}}
		& Macro-Sensitivity & 56.19 $\pm$ 1.37 & 28.69 $\pm$ 6.93 & 58.44 $\pm$ 5.73 & \textbf{33.69 $\pm$ 5.09} & \textbf{60.91 $\pm$ 2.49} & 30.42 $\pm$ 3.49 \\
		& Macro-Specificity & 86.31 $\pm$ 0.59 & 79.16 $\pm$ 2.06 & 87.46 $\pm$ 2.33 & \textbf{80.27 $\pm$ 1.56} & \textbf{88.00 $\pm$ 1.19} & 80.25 $\pm$ 0.78 \\
		& Macro-PPV         & 51.29 $\pm$ 10.40 & 30.93 $\pm$ 23.72 & 49.33 $\pm$ 5.65 & \textbf{35.24 $\pm$ 11.56} & \textbf{51.80 $\pm$ 2.65} & 21.24 $\pm$ 1.59 \\
		& Macro-Accuracy    & 71.25 $\pm$ 0.89 & 53.92 $\pm$ 4.48 & 72.95 $\pm$ 4.03 & \textbf{56.98 $\pm$ 3.31} & \textbf{74.45 $\pm$ 1.84} & 55.33 $\pm$ 2.08 \\
		& Macro-F1-score    & 51.55 $\pm$ 3.50 & 24.30 $\pm$ 9.23 & 52.82 $\pm$ 5.84 & \textbf{27.66 $\pm$ 4.29} & \textbf{54.39 $\pm$ 3.11} & 23.76 $\pm$ 1.77 \\
		& Micro-Accuracy    & 80.67 $\pm$ 0.91 & 70.00 $\pm$ 2.89 & 82.33 $\pm$ 3.46 & \textbf{71.33 $\pm$ 2.17} & \textbf{83.00 $\pm$ 1.83} & 69.00 $\pm$ 1.49 \\
		
		\hline
	\end{tabular}
	\label{Table_Metrics}
\end{table*}

\subsection{Agreement metrics}

\begin{figure*}[!t]
	\centering 
	\includegraphics[width=15.5cm]{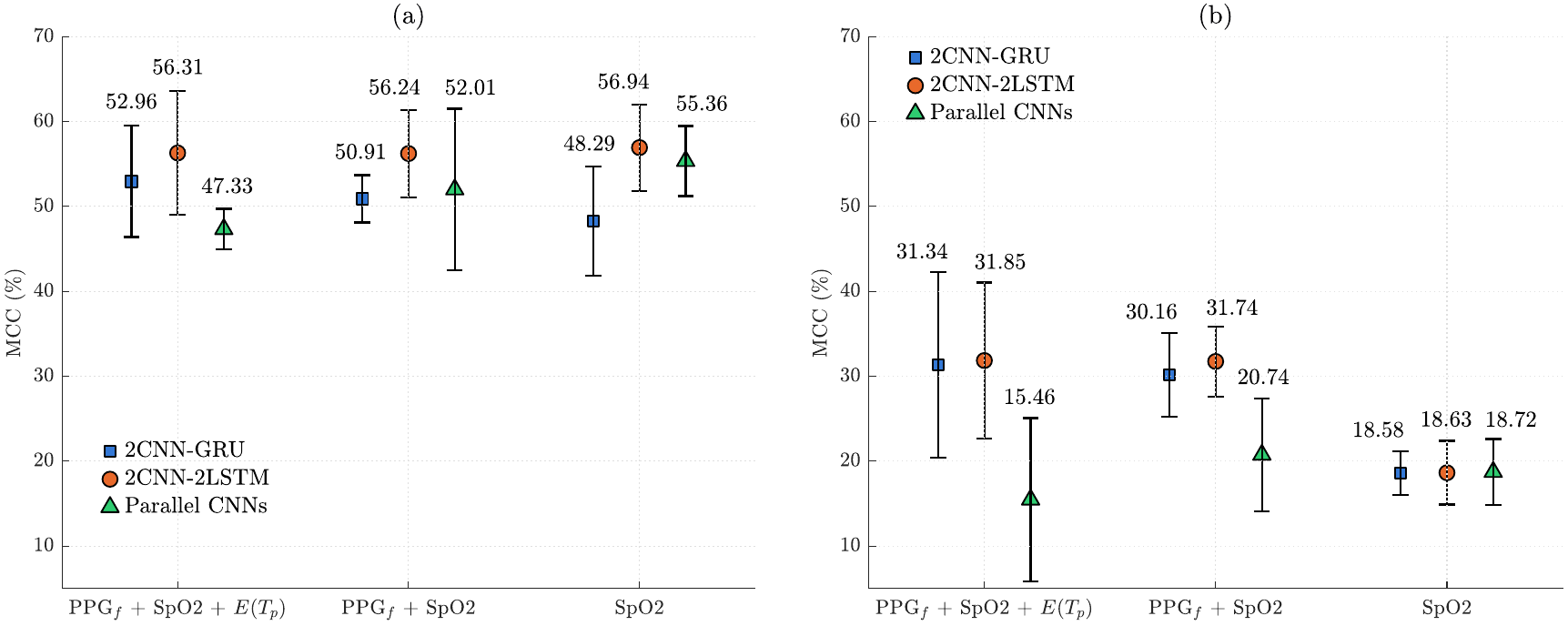} 
	\caption{Matthews correlation coefficient, MCC, of classifying sleep apnea patients when using different model architectures and predictor combinations for (a) ID testing and (b) OOD testing data. Error bars (mean $\pm$ standard deviation) obtained from five iterations when different fixed random seeds were used. PPG$_f$ corresponds to five pulse wave features, whereas $E(T_p)$ -- the mean envelope feature of the pulse wave interval. Mean values are shown above error bars.}
	\label{Fig20}
\end{figure*}

\begin{figure*}[!t]
	\centering 
	\includegraphics[width=15.5cm]{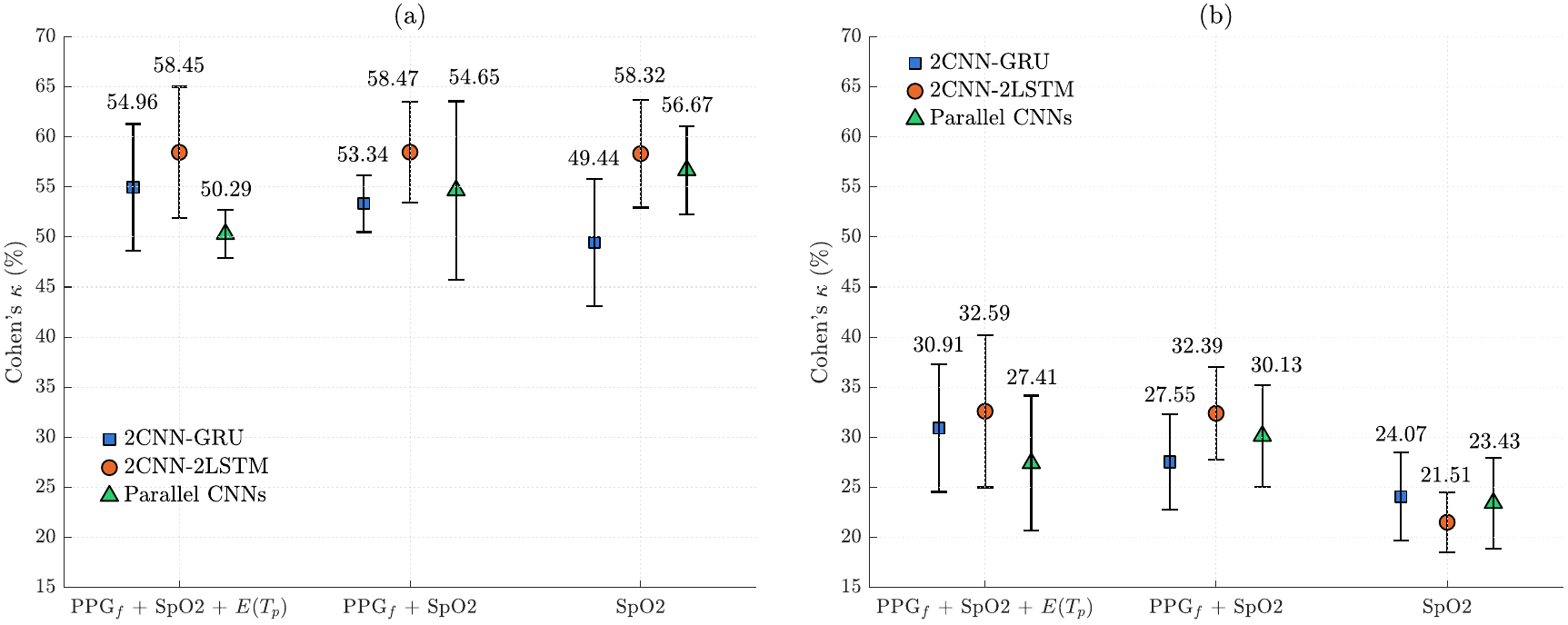} 
	\caption{Cohen’s kappa of classifying sleep apnea patients when using different model architectures and predictor combinations for (a) ID testing and (b) OOD testing data. Error bars (mean $\pm$ standard deviation) obtained from five iterations when different fixed random seeds were used. PPG$_f$ corresponds to five pulse wave features, whereas $E(T_p)$ -- the mean envelope feature of the pulse wave interval. Mean values are shown above error bars.}
	\label{Fig21}
\end{figure*}

The agreement metrics calculated between predicted and true AHI values are shown in Figures~\textcolor{blue}{\ref{Fig20}}~and~\textcolor{blue}{\ref{Fig21}}. The highest OOD Cohen's kappa was obtained by the 2CNN-2LSTM model using PPG pulse wave features, SpO2, and $E(T_p)$, reaching 32.59~$\pm$~7.59\%, while the highest OOD MCC was achieved by the same model and predictor combination, reaching 31.85~$\pm$~9.20\%. The 2CNN-GRU model performed slightly worse, with OOD Cohen's kappa and MCC values of 30.91~$\pm$~6.36\% and 31.34~$\pm$~10.93\%, respectively, when PPG pulse wave features, SpO2, and $E(T_p)$ were used. The weakest OOD agreement was observed for the parallel CNNs architecture, when MCC dropped to 15.46~$\pm$~9.60\%.

\section{Discussion}
\label{Section4}
This study proposed the novel AB prediction-based framework for characterizing sleep apnea severity and investigated the impact of PPG features on model performance and OOD generalization. The obtained results support the hypothesis that combining PPG-derived features with SpO2 enhances classification performance, particularly under the evaluated OOD scenarios.

\subsection{Method Novelty $\&$ Potential Advantages}
The method novelty of this study lies in the use of regression-based model architectures for sleep apnea characterization, in contrast to the majority of existing approaches that rely on direct classification frameworks\textcolor{blue}{\hyperlink{ref9}{\cite{ref9}}},\textcolor{blue}{\hyperlink{ref14c}{\cite{ref14c}}},\textcolor{blue}{\hyperlink{ref15}{\cite{ref15}}},\textcolor{blue}{\hyperlink{ref15b}{\cite{ref15b}}},\textcolor{blue}{\hyperlink{ref15c}{\cite{ref15c}}},\textcolor{blue}{\hyperlink{ref15d}{\cite{ref15d}}},\textcolor{blue}{\hyperlink{ref30}{\cite{ref30}}}. Specifically, the proposed models employ a regression output layer to estimate AB, which is subsequently transformed into the clinically relevant AHI metric (see Figure~\textcolor{blue}{\ref{Fig2}} and Equation~\textcolor{blue}{\ref{AHI_AB}}). This formulation enables a two-stage interpretation: continuous estimation followed by classification-per-subject. An important advantage of this approach is its reduced dependence on predefined decision thresholds, which are typically required in classification-based models and may introduce bias or sensitivity to class imbalance. Furthermore, predicting a continuous physiological quantity provides a more flexible and clinically meaningful representation of disease severity, enabling post~hoc categorization according to established AHI thresholds. As a result, the proposed framework better aligns with current clinical practice, where subject-level assessment of sleep apnea is essential for diagnosis, monitoring, and treatment planning.

From a clinical perspective, the proposed framework could be integrated into sleep apnea screening workflows as a decision-support tool for subject-level severity assessment based on overnight PPG and SpO2 recordings. Owing to the non-invasive nature and widespread availability of these signals in wearable devices, it could facilitate longitudinal monitoring of disease severity and treatment response, for example during continuous positive airway pressure therapy. Furthermore, because the proposed framework relies only on PPG and SpO2 analysis rather than full PSG, it may provide a more accessible and cost-effective solution for long-term follow-up in both clinical and home settings.

Although the primary focus of this study is on OOD generalization, the obtained ID performance results are also of significant practical importance. High ID accuracy indicates that the proposed models are capable of capturing subject group-specific physiological patterns, which is particularly relevant in personalized or patient-adaptive systems as proposed in\textcolor{blue}{\hyperlink{ref31}{\cite{ref31}}}. In such scenarios, periodic model recalibration using newly acquired patient data could further improve performance over time. This adaptive approach may reduce inter-subject variability and enhance prediction reliability in longitudinal monitoring settings. 

\subsection{Model Architectures $\&$ Features} 
Within the evaluated experimental setting, the 2CNN-2LSTM architecture achieved the most favorable balance between classification performance and OOD robustness. A comparison of predictor combinations showed that adding PPG-derived features generally can improve OOD generalization for low-complexity hybrid architectures. In the case of both 2CNN-GRU and 2CNN-2LSTM, PPG-included predictor combinations outperformed SpO2 alone across OOD performance and agreement metrics. In contrast, for the parallel CNNs model, the best ID results were obtained using only SpO2.

Since each architecture was individually optimized, the reported comparison should be interpreted as exploratory rather than as a definitive benchmark of architectural superiority. The primary objective of this study was to evaluate the proposed AB-guided framework and the contribution of PPG-derived features, rather than to identify the optimal neural network architecture.

The most substantial OOD generalization improvements were observed for the hybrid convolutional-recurrent architectures, particularly the 2CNN-2LSTM model, which achieved the highest gains in macro-F1-score (+16.01\%), Cohen's kappa (+11.08\%), and MCC (+13.22\%). Similarly, the 2CNN-GRU model exhibited the strongest improvement in macro-sensitivity (+16.07\%) and macro-PPV (+19.60\%), indicating enhanced detection capability for apnea severity classes under distribution shift. While the parallel CNNs architecture showed notably smaller improvements across most metrics (the largest increases were +14.00\% and +6.70\% in PPV and Cohen's kappa, respectively), suggesting a lower benefit from additional PPG features in this multi-scale convolutional-only design. Therefore, these findings suggest that, within the scope of the evaluated models, datasets, and training setup, PPG features tend to provide the greatest benefit when combined with low-complexity hybrid architectures, leading to improved robustness in OOD scenarios.

Regarding the ID testing, the contribution of PPG features was relatively modest, which is consistent with findings reported in\textcolor{blue}{\hyperlink{ref31b}{\cite{ref31b}}}, where accuracy of detecting apnea events reached 68.70\%, 82.20\%, and 83.40\% when using PPG features, SpO2, and their combined analysis, respectively.

From a practical perspective, our study supports using 2CNN-2LSTM and 2CNN-GRU models for sleep apnea characterization. In addition to the conventional SpO2 predictor, the inclusion of PPG-derived biomarkers appears valuable, especially when robustness to distribution shifts is required in real-world applications.

\subsection{Limitations}

Despite the promising results, several limitations should be acknowledged. The evaluation was conducted on a limited number of datasets, which may not fully represent real-world clinical variability. A noticeable performance gap between ID and OOD results remains, indicating sensitivity to distribution shifts. In some cases including the proposed feature $E(T_p)$ exhibited increased variability across training runs, suggesting limited performance stability.

Although the external OSASUD dataset provided an opportunity to evaluate the proposed framework under a clinically meaningful domain shift, it includes only 30 subjects. While this cohort differs substantially from the training population in terms of clinical status, acquisition conditions, and recording equipment (see Table~\textcolor{blue}{\ref{Table1}}), its limited size may affect the generalizability of the OOD results. Therefore, the presented OOD performance should be interpreted as preliminary evidence supporting the findings of the proposed framework to an independent external cohort.

Another important limitation is related to the estimation of the AHI. The predicted AHI was derived from the predicted AB using a fixed mean event duration $D_{\mathrm{mean}}$, calculated from the training data. However, in real-world conditions, the duration of apnea/hypopnea events is highly variable and can range approximately from 10 seconds to up to 2 minutes. Therefore, the use of a fixed average duration may introduce estimation bias and limit the accuracy of the AHI, particularly across different populations and sleep conditions. Due to potential conversion errors (see Section~\textcolor{blue}{\ref{SeverityAssessment}} and Figure~\textcolor{blue}{\ref{Fig2}}), borderline severity cases should be examined more thoroughly. Nevertheless, this approach was adopted as a practical simplification to enable a consistent transformation from AB to a clinically interpretable metric within the proposed framework.

\subsection{Future Work}

Further studies should evaluate the proposed framework using larger and more diverse external datasets to strengthen the evidence for its generalizability. Future work will also focus on improving model robustness and interpretability, including deeper analysis of feature contributions and behavior. For instance, additional PPG-related biomarkers could be investigated, as we did with the derivative-based feature, $S_{max}$, to further enhance model performance.

Moreover, the combination of PPG features and SpO2 could be explored using alternative deep learning architectures, such as transformer-based models, which have recently shown strong potential for modeling long-range dependencies in physiological signals\textcolor{blue}{\hyperlink{ref12}{\cite{ref12}}},\textcolor{blue}{\hyperlink{ref14}{\cite{ref14}}},\textcolor{blue}{\hyperlink{ref14b}{\cite{ref14b}}},\textcolor{blue}{\hyperlink{ref14c}{\cite{ref14c}}}. Since the proposed AB-guided framework is model-agnostic and independent of the underlying learning paradigm, it can also be integrated with self-supervised and foundation-model approaches. Although the evaluation of these approaches was beyond the scope of the present study, they represent promising directions for future research.

In addition, advanced techniques such as domain adaptation\textcolor{blue}{\hyperlink{ref32}{\cite{ref32}}} and transfer learning could be explored to improve generalization under distribution variance. Another promising direction is the development of adaptive, patient-specific models with periodic recalibration.

Future research could investigate the determination of clinically meaningful AB cut-off thresholds for direct classification into severity categories. Such an approach would eliminate the need for transformation from AB to AHI and could simplify the overall framework while preserving clinical interpretability. In addition, future studies should investigate patient-specific AB-to-AHI conversion strategies and perform dedicated sensitivity or error propagation analyses to further quantify the contribution of the conversion model to the overall prediction error.

\section{Conclusion}
\label{Section5}

We hypothesized that combining PPG pulse wave features with SpO2 in artificial neural networks would improve sleep apnea severity classification and enhance OOD generalization. Our results support this hypothesis: fusing PPG-derived features with SpO2 and employing hybrid convolutional-recurrent architectures improved classification performance, particularly under the evaluated distribution shifts. Among the implemented models, the 2CNN-2LSTM architecture achieved the highest overall performance.

The findings highlight the importance of jointly optimizing model architecture and feature representation for robust and generalizable sleep apnea assessment. The new proposed approach provides a clinically relevant AB-guided framework for subject-level severity estimation and has strong potential for deployment in non-invasive, real-world, and home-based sleep monitoring applications.

\section*{Declaration of competing interest}
The authors declare that they have no known competing financial interests or personal relationships that could have influenced the work reported in this manuscript.

\section*{Data availability}
The MATLAB scripts for preparing and partitioning the MESA and OSASUD datasets are available \textcolor{blue}{\href{https://gitlab.com/qumphy/d4-code}{online}}.

\begin{IEEEbiographynophoto}{Mantas Rinkevi\v{c}ius} received the B.Sc. and M.Sc. degrees in biomedical engineering from Kaunas University of Technology (KTU), Lithuania, in 2019 and 2021, respectively, and the Ph.D. degree in electrical and electronics engineering from KTU, Lithuania, in 2025.
	
From 2019 to 2023, he was a Project Engineer with the Biomedical Engineering Institute, KTU, and UAB Gruppo Fos
Lithuania. From 2023 to 2025, he was a Junior Researcher with the Biomedical Engineering Institute, KTU. Since 2026, he has been a Scientific Researcher with the Biomedical Engineering Institute, KTU, Lithuania. His research interests include sleep apnea detection and characterization, photoplethysmogram signal quality analysis, cuffless blood pressure estimation, autonomic nervous system assessment, stress monitoring, heart rate variability and arrhythmia analysis, biomedical signal processing and feature extraction, and machine learning and deep learning methods.
	
Dr. Rinkevičius received the KTU Talent Scholarship from 2017 to 2021, the Engineer Vytautas Naginevičius Scholarship in 2018, the President of the Republic of Lithuania Jonas Žemaitis Scholarship in 2020--2021, the Award for the Most Active KTU Ph.D. Student in 2024, and the Research Council of Lithuania Scholarship for Ph.D Studies in 2025.
\end{IEEEbiographynophoto}

\begin{IEEEbiographynophoto}{Oskar Pfeffer} received the M.Sc. degree in theoretical physics from the Technical University of Berlin in 2021. Since 2023, he has been a PhD candidate at the Physikalisch-Technische Bundesanstalt in Berlin.
From 2019 to 2021, he worked as a Student Assistant at the Potsdam Institute for Climate Impact Research, focusing on complex energy networks. His current research spans uncertainty quantification in machine learning models for medical applications, as well as quantum algorithms for partial differential equations and tensor networks.
\end{IEEEbiographynophoto}

\begin{IEEEbiographynophoto}{Amal Alissa} received the B.Sc. degree in mathematics from the Lebanese University, Lebanon, in 2021. She is currently pursuing the M.Sc. degree in mathematics at Freie Universität Berlin, Germany. Alongside the studies, she is working as a Research Assistant at the Physikalisch-Technische Bundesanstalt, Department for Mathematical Modelling and Data Analysis. Her current work focuses on machine learning and its applications in healthcare.
\end{IEEEbiographynophoto}

\begin{IEEEbiographynophoto}{Nando Hegemann} received the M.Sc. degree in mathematics from the Humboldt Universität zu Berlin in 2017 and the Ph.D. degree in mathematics from the Technical University Berlin in 2022. Since 2021, he has been a Senior Research Scientist at the Physikalisch-Technische Bundesanstalt, Department for Mathematical Modelling and Data Analysis. He has expertise in mathematical and computational modelling of inverse problems and uncertainty quantification. Recently, his work focusses on uncertainty quantification of machine learning and deep learning for optical and medical applications.
\end{IEEEbiographynophoto}

\vspace{-175pt}
\begin{IEEEbiographynophoto}{Vaidotas Marozas} (Member, IEEE) received the B.Sc., M.Sc., and Ph.D. degrees in electronics engineering from Kaunas University of Technology (KTU), Lithuania, in 1993, 1995, and 2000, respectively. He is currently a Professor with the Department of Electronics Engineering and the Director of the Biomedical Engineering Institute at KTU.

His research interests include biomedical signal processing, physiological modeling, digital biomarkers, wearable systems for health monitoring, cardiovascular disease assessment, and machine learning methods for biomedical applications. He has led and participated in numerous national and international research projects, including those funded by Horizon Europe, FP7, H2020, COST Action, and the Swedish Institute Visby Programme, and has coordinated multiple R$\&$D projects with industrial partners.

Prof. Dr. Marozas has authored and co-authored numerous scientific publications in leading journals, including IEEE Transactions on Biomedical Engineering and the IEEE Journal of Biomedical and Health Informatics, and is a co-author of monographs published by Springer and Academic Press. He has supervised seven Ph.D. students and has received several awards, including KTU Researcher of the Year (2016) and KTU Project Manager of the Year (2025).
\end{IEEEbiographynophoto}

\EOD

\end{document}